%% file: paper_arxiv_clean.tex
\RequirePackage{snapshot}
 \documentclass[12pt]{article}
 \usepackage{amsmath}
 \usepackage{amssymb}
 \usepackage{bm}
 \usepackage[]{color}
 \usepackage{natbib}

\definecolor{fgcolor}{rgb}{0.345, 0.345, 0.345} 
\definecolor{shadecolor}{rgb}{.97, .97, .97}
\definecolor{messagecolor}{rgb}{0, 0, 0}
\definecolor{warningcolor}{rgb}{1, 0, 1}
\definecolor{errorcolor}{rgb}{1, 0, 0}

\usepackage{dsfont}
\usepackage{graphicx}
\usepackage[usenames,dvipsnames,svgnames,table]{xcolor}

\usepackage[margin=1in]{geometry}
\usepackage{setspace}
\title{Multiple Imputation of Missing Categorical and Continuous Values via Bayesian Mixture Models with Local Dependence} 
\author{Jared S. Murray 
 and Jerome P. Reiter
 \footnote{Jared S. Murray (jsmurray@stat.cmu.edu) is Visiting
   Assistant Professor, Department of Statistics, Carnegie Mellon
   University. Jerome P. Reiter is the Mrs. Alexander Hehmeyer
   Professor of Statistical 
Science, Duke University. This work was supported by grants from the National Science Foundation (SES-11-31897, SES-1130706 and DMS-1043903). Any opinions, findings, and conclusions or recommendations  
expressed in this material are those of the author(s) and do not necessarily
 reflect the views of the Census Bureau or National Science Foundation. }
}
\date{\today}
\newcommand{\distas}[1]{\mathbin{\overset{#1}{\kern\z@\sim}}}%
\newcommand{\ind}[1]{\mathds{1}(#1)}%
\newsavebox{\mybox}\newsavebox{\mysim}
\newcommand{\iid}{\overset{iid}{\sim}}

\newcommand{\ac}{HCMM-LD}

\DeclareMathOperator{\logit}{logit}

\newcommand{\Hy}{H^{(\mathcal{Y})}}
\newcommand{\Hx}{H^{(\mathcal{X})}}
\newcommand{\Hyi}{H_i^{(\mathcal{Y})}}
\newcommand{\Hxi}{H_i^{(\mathcal{X})}}
\newcommand{\hy}{r}%
\newcommand{\hx}{s}%
\newcommand{\ky}{{k^{\mathcal{(Y)}}}}
\newcommand{\kx}{{k^{\mathcal{(X)}}}}
\newcommand{\kz}{{k^{\mathcal{(Z)}}}}
\newcommand{\byx}[1]{#1^{(\mathcal{X})}}
\newcommand{\byy}[1]{#1^{(\mathcal{Y})}}
\newcommand{\byz}[1]{#1^{(\mathcal{Z})}}
\newcommand{\zmax}{{k^{\mathcal{(Z)}}}}

\newcommand\id{\perp\!\!\!\perp}

\DeclareMathOperator*{\diag}{diag}
\begin{document}

\maketitle

\begin{abstract}
We present a nonparametric Bayesian joint model for multivariate continuous and categorical variables, with the intention of developing a flexible engine for multiple imputation of missing values.   
The model fuses Dirichlet process mixtures of multinomial
distributions for categorical variables with Dirichlet process mixtures of multivariate normal distributions for continuous variables.  We incorporate dependence between 
the continuous and categorical variables by (i) modeling the means of the normal distributions as component-specific functions of the categorical variables and (ii) forming distinct mixture components for 
the categorical and continuous data with probabilities that are linked via a hierarchical model. This structure allows  the model to capture complex dependencies between the categorical and continuous data with 
minimal tuning by the analyst. 
We apply the model to impute missing values due to item nonresponse in an evaluation of the redesign of the Survey of Income and Program Participation (SIPP). %
The goal is to compare estimates from a field test with the new design to estimates from selected individuals from a panel collected under the old design.  We
show that accounting for the missing data changes some conclusions about the comparability of the distributions in the two datasets.  We also perform an extensive repeated sampling simulation using similar data from 
complete cases in an existing SIPP panel, comparing our proposed model to a default application of multiple imputation by chained equations.  Imputations based on the proposed model tend to have better repeated sampling properties than 
the default application of chained equations in this realistic setting.
\end{abstract}

\section{Introduction}

The Survey of Income and Program Participation (SIPP) is the largest government survey of people 
on public assistance in the United States. It includes longitudinal 
data on income, labor force information, participation and eligibility
for governmental assistance programs, and general demographic
characteristics for individuals on public assistance; as such, it is used by a broad community of researchers and policy-makers  \citep{kinnerreiterjos}.
In 2014, the Census Bureau redesigned the SIPP to utilize a longer
reference period (twelve months, instead of four) and a new instrument
that incorporates an event  history calendar  \citep{Moore2009}. The
Census Bureau made these changes with the hope of reducing costs and
respondent burden while improving accuracy.

To evaluate the redesign,  the Census Bureau conducted a field test by giving the new survey to a non-overlapping sample of individuals drawn from the 
same frame used to construct the 2008 production SIPP panel. The field test was restricted to individuals in low income strata in 20 states.
The Census  Bureau also constructed a comparison dataset from the production SIPP panel comprising individuals from the same strata and states, with the intention of  
assessing whether or not the change in collection instruments resulted in different distributions of key variables. Additional details 
are available in \cite{USCensus2013}.

The data from the field test suffered from item nonresponse, as did the data
from the production SIPP.  For example, among the sampled field 
test individuals, approximately 16\% are missing employment status and, for those who reported participation in the Supplemental Nutrition Assistance Program (SNAP),
approximately 59\% are missing the benefit amounts.  Unless the
missing data mechanisms are identical in both datasets, 
e.g., both missing completely at random \citep{Rubin1976},
comparisons of available case analyses may result in inaccurate
conclusions about where estimates from the old and new designs differ. 

Given the objective of comparing two datasets on many
analyses, a sensible approach is to create and utilize
multiply-imputed versions \citep{Rubin1987} of each sample.  In
multiple imputation (MI) the imputer repeatedly samples values of the
missing data from their predictive distribution under an appropriate
 model to create $m>1$ completed datasets.  The analyst
then computes point and variance estimates in each of the $m$
datasets, and combines them using straightforward rules 
\citep{Rubin1987, reiter:raghu:07}.  These rules allow the analyst to account for
uncertainty due to the missing data when making inferences. 

The SIPP data have distributional features that are
challenging to capture with imputation based on
standard (semi-)parametric models. 
For example, some continuous variables have different variances and skewness at
different combinations of the categorical variables, and the
categorical variables have complex dependencies.  
Thus, it is desirable to use imputation models that can capture such features in
each dataset with minimal tuning.

In this article, we introduce a nonparametric Bayesian joint
model for mixed continuous and categorical data suitable for use as a
flexible, fully coherent multiple imputation engine.  The basic idea
is to 
fuse two Dirichlet Process (DP) mixtures: a mixture of multinomial distributions (for the categorical data) and a mixture of multivariate normal regressions (for the continuous data, conditionally on the categorical variables).
We model dependence between the
categorical and  continuous variables by (i) 
specifying the means of the normal distributions
as component-specific functions of the categorical variables, and (ii) inducing dependence in the separate component assignments via a
hierarchical model. As we illustrate, the model includes
local dependence---i.e., dependence among variables within mixture components---between the categorical and continuous data; 
thus, we call it a hierarchically coupled mixture model with local
dependence (\ac). Local dependence allows the model to more efficiently capture complex dependence structure in observed variables.

This ability to capture complex dependence, as well
as conform to different distributional shapes, is attractive for
multiple imputation contexts, as it  helps the imputer to preserve
structure in the data that he or she may not have anticipated but may
be important to analysts.  Mixture models have been suggested previously for multiple imputation of missing 
categorical data \citep[e.g.,][]{Vermunt2008, Gebregziabher2010, Si2013, 
  Manrique-Vallier2012, Manrique-Vallier2012a, si:reiter:hillygus:14} and missing continuous data \citep[e.g.,][]{Bohning2007,
  Elliott2007, Kim2013}. To our knowledge, mixture models have not been used to impute mixed categorical and continuous data.

The remainder of the article proceeds as follows.  In Section
\ref{sec:motivation},  we illustrate some of the complex
distributional features of the variables in the SIPP, and we discuss
how existing multiple imputation routines could struggle to capture
such features. In Section \ref{sec:model}, we introduce the \ac\
including specification of prior distributions, and discuss related
nonparametric Bayesian models. %
In Section \ref{sec:sipp-sim}, we present results of a repeated sampling
simulation using complete cases from an existing SIPP panel, 
illustrating the potential for improved performance of multiple imputation using the
\ac\ over a default application of MI by chained
equations \citep{Raghunathan2001,VanBuuren1999}. In Section
\ref{sec:msipp}, we apply the \ac\ to multiply-impute missing values
in the field test data as well as the representative subsample of the production SIPP. Some
conclusions about the comparability of the two designs change
after accounting for the missing data.  Finally, in Section
\ref{sec:conclusion}, we conclude with a discussion of extensions and
future work.

\section{The Challenges of Imputing SIPP Data}\label{sec:motivation}

\begin{figure}
 \centering
 \includegraphics[width=.9\textwidth]{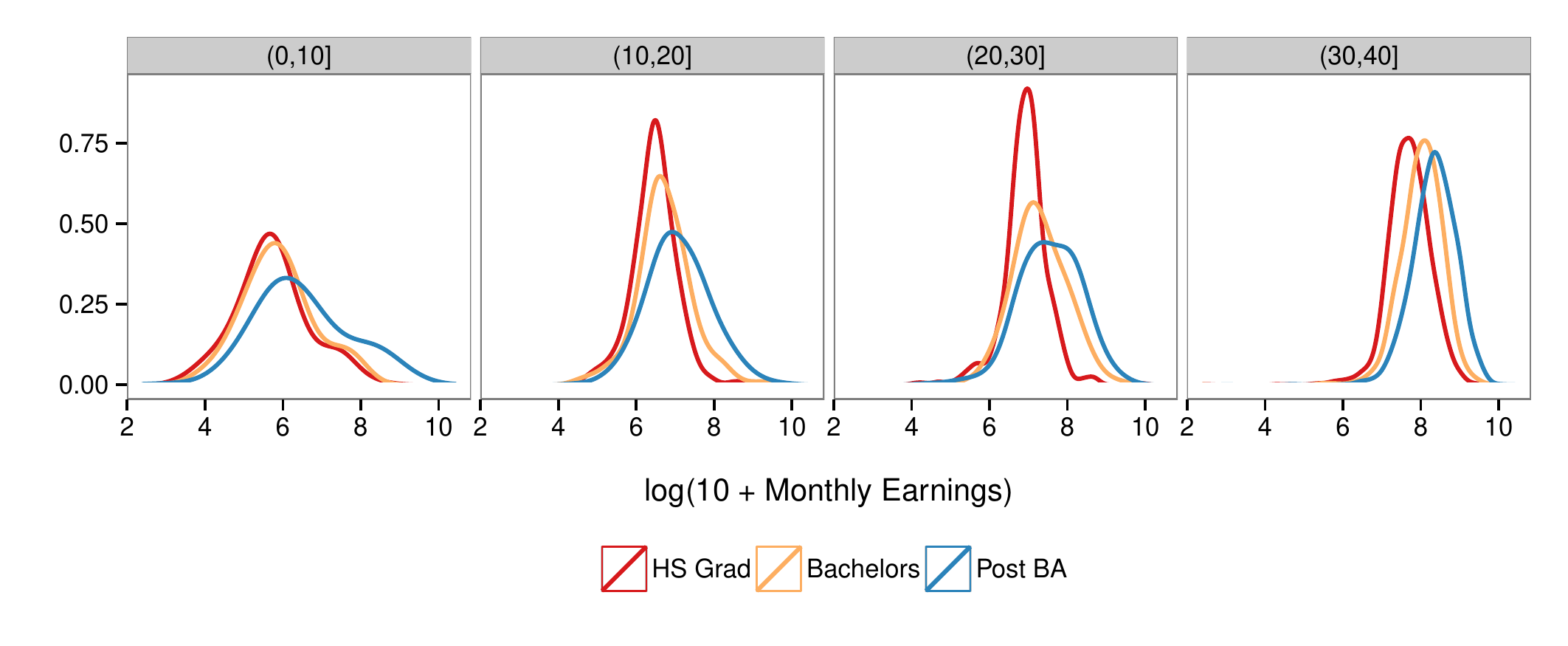}
 \caption{Log monthly earnings from SIPP, by usual hours worked and education level}
 \label{fig:earn_by_hrs_edu}
\end{figure}

SIPP is characteristic of many surveys in that it includes many
categorical variables and a smaller number of continuous variables, with
complicated dependence and nonstandard distributions. For example, using public-use data from the 2008 SIPP panel, Figure \ref{fig:earn_by_hrs_edu} displays
plots of (log) total earnings by usual hours worked and education level. The
distribution of income varies across levels of the discrete variables.
 The earnings distribution is skewed right when usual hours $\leq 10$, whereas it
eventually becomes slightly skewed left as the number of hours increases. In the
 first three panels, increasing education level is associated with increased
 dispersion in the distribution of log earnings. In the last panel, increased
 education is primarily associated with a location shift in earnings. 
There is also evidence of higher-order dependence in the distributions
of the categorical variables. Table \ref{tab:loglm} shows analysis of deviance
tables for one, two, and three way loglinear models fit to a few
subsets of the categorical variables. All indicate some evidence of
interactions.

\begin{table}[t]
\caption{Analysis of deviance tables for loglinear models fit to subsets of the SIPP data.}\label{tab:loglm}
\centering
\begin{tabular}{lrrrrr}

 \multicolumn{6}{c}{Race, sex, education level, hourly}\\
  \hline
 & Resid. Df & Resid. Dev & Df & Deviance & Pr($>$Chi) \\ 
  \hline
\multicolumn{6}{l}{Race, sex, education level, hourly}\\
$\,\,\,$1 way & 108 & 10191.25 &  &  &  \\ 
$\,\,\,$  2 way & 69 & 211.50 & 39 & 9979.76 & $<10^{-6}$ \\ 
$\,\,\,$  3 way & 20 & 39.92 & 49 & 171.58 & $<10^{-6}$ \\ 
 &  & &  &  &  \\ 
 \multicolumn{6}{l}{Marital status, usual hours, sex, no. own children}\\
$\,\,\,$1 way & 132 & 6350.60 &  &  &  \\ 
$\,\,\,$  2 way & 91 & 1383.23 & 41 & 4967.36 & $<10^{-6}$ \\ 
$\,\,\,$  3 way & 30 & 168.83 & 61 & 1214.40 & $<10^{-6}$ \\ 
 \hline

\end{tabular}
\end{table}

Given these complex distributional features, what sort of models might
one use for imputation of missing values?  One possible approach is to use
a general location model (GLOM) \citep{Olkin1961,Little1985,Schafer1997}. For
continuous variables $Y$ and discrete  variables $X$,  the GLOM assumes that $(Y\mid X=x)\sim
N(\mu_x, \Sigma_x)$ and $X\sim \pi$ with $\pi\sim Dir(\alpha)$; see
also \cite{Liu1998} who generalize the $(Y\mid X)$ model to the class
of elliptically symmetric distributions. Estimation under this model
is infeasible unless each cell of the table implied by $X$ contains a large number of
observations.  Thus, imputers typically impose further constraints,
most often that $\Sigma_x\equiv \Sigma$ for all $x$,
$\mu_x = D(x)B$ for a matrix of regression coefficients $B$ and design
vector $D(x)$, and $\pi$ satisfies loglinear constraints that include
interactions only up to a certain order.  Multivariate
normality and common covariance structure seem unlikely to fit the
types of features apparent in Figure \ref{fig:earn_by_hrs_edu}.
Further, Table \ref{tab:loglm} suggests that it would be easy to 
miss key interactions when selecting the loglinear model.  Thus,
the GLOM seems overly restrictive for the SIPP data.

An alternative approach  is to specify a sequence of univariate models 
for each variable subject to missingness conditional on subsets of
the other variables, e.g., impute $a$ from $f(a \mid b,c)$, impute $b$
from $f(b \mid a,c)$, and impute $c$ from $f(c \mid a,b)$. This is
known as the ``chained equations'' or ``fully-conditional'' approach
\citep{Raghunathan2001,VanBuuren1999}. 
While multiple imputation by chained equations (MICE) approaches have proven to be quite useful for
many datasets, they can be challenging to use
effectively for data with complex dependence like the
SIPP.  For example, typical applications of MICE use
multinomial logistic regressions for the categorical
variables.  Relationships between an outcome and the
remaining predictors may be nonlinear and involve interaction effects;
these can be difficult to find when the data have more than a handful
of variables (that may also be subject to missingness). Similar challenges arise when specifying models for continuous data, even with semiparametric extensions like predictive mean matching \citep{Little1988}.
Additionally, %
the selected conditional models may be incompatible; that is, there may not be any joint model with the specified conditionals \citep{Liu2014}. 
This may result in imputation procedures with undesirable theoretical properties \citep{Si2013}.

A third and related approach is to specify a coherent joint distribution as a
sequence of conditional models \citep{ibrahimlips, ibrahim, ibrahim:chen:lip:herr}, for example $f(a, b,
c) = f(a) f(b \mid a) f(c \mid b, a)$.  
Compared to typical chained equations
approaches, this has the advantage of resulting in a fully coherent joint 
distribution.  However, it still can be challenging to find and model complex
distributional features, particularly for models with many predictors.  Additionally, 
different conditioning sequences for the variables could result in different fits, 
and the imputer may not have good information to help choose an order.

\section{Multiple Imputation via the HCMM-LD}\label{sec:model}

For $i=1,\dots,n$ sampled individuals, let
$X_i = (X_{i1},\dots, X_{ip})'$ be a vector of $p$  
categorical variables for individual $i$, with each $X_{ij} \in
\{1,\dots, d_j\}$, and let $Y_i = (Y_{i1},\dots, Y_{iq})'$ be a vector of $q$ continuous responses taking values in $\mathbb{R}^q$. 
We use $x_i$ and $y_i$ for specific values taken by $X_i$ and $Y_i$.
We also use superscript $\mathcal{X}$ and $\mathcal{Y}$ to signify that some parameter or latent variable is a component of the 
model for $X$ or $Y$, respectively.

\subsection{The HCMM-LD for Imputing Mixed Data}

As noted in Section 1, mixture models have proven valuable for imputing multivariate missing data that are 
strictly continuous or categorical.  
The \ac\ fuses existing mixture models for strictly continuous or categorical data into a larger hierarchical model.
Thus, we begin with a brief 
summary of these existing mixture models and discuss the shortcomings of various ``intuitive'' ways to combine them. We  
present the \ac\ in Section \ref{sec:indexmodel}.

For imputing multivariate continuous data, \cite{Kim2013} use a truncated Dirichlet process (DP) mixture of normal distributions.
For $i=1,\dots,n$, let $\Hyi \in \{1, \dots, \ky\}$ be the mixture component index for record $i$. This model assumes that 
\begin{equation}
(Y_i\mid \Hyi=\hy, \{(\mu_\hy, \Sigma_\hy) : 1\leq \hy\leq \ky\}) \sim N(\mu_\hy, \Sigma_\hy).\label{eq:ymodel}
\end{equation}
The prior distribution for $\Hyi$ is a truncated version of the stick-breaking construction for the DP \citep{Sethuraman1994}, introduced in \cite{Ishwaran2001}:
\begin{gather}
\Pr(\Hyi=\hy) = \byy{\phi_{\hy}}\\
\byy{\phi_{\hy}} = \byy{\xi_\hy}\prod_{l<\hy}(1-\byy{\xi_l}),\ \{\byy{\xi_\hy}: 1\leq \hy\leq \ky-1 \}\iid Beta(1, \byy{\beta}),\ 
\byy{\xi_{\ky}}\equiv 1.\label{eq:Kip}
\end{gather}

For imputing multivariate categorical data, \cite{Si2013} adopt a truncated version of the DP mixture of product multinomials (MPMN) proposed
by \cite{Dunson2009}. 
For $i=1,\dots,n$, let $\Hxi \in \{1, \dots, \kx\}$ be the mixture component index for record $i$, 
and let $\Pr(X_{ij} = x_{ij} \mid \Hxi = \hx) = \psi_{\hx x_{ij}}^{(j)}$. This model assumes that  
\begin{align}
 \Pr(X_i = x_i \mid \Hxi = \hx, \{\psi_\hx : 1\leq \hx\leq\kx\}) &= \prod_{j=1}^{p}\psi_{\hx x_{ij}}^{(j)}\label{eq:dx},
\end{align}
where, for each $1\leq \hx\leq \kx$,  $\psi_\hx=\{ \psi_{\hx}^{(j)}: 1\leq j \leq p\}$ and each 
$\psi^{(j)}_\hx = (\psi^{(j)}_{\hx 1}, \dots, \psi^{(j)}_{\hx d_j})'$ is a probability vector.
The prior on $\Pr(\Hxi=\hx)$ is another truncated stick breaking process:
\begin{gather}
\Pr(\Hxi=\hx) = \byx{\phi_\hx}\\
\byx{\phi_\hx} = \byx{\xi_\hx}\prod_{l<\hx}(1-\byx{\xi_l}),\ \{\byx{\xi_\hx}: 1\leq \hx\leq \kx-1 \}\iid Beta(1, \byx{\beta}),\ \byx{\xi_\kx}\equiv 1.\label{eq:Hip}
\end{gather}

Given their success as imputation engines, it seems promising to fuse these two models into a coherent joint 
distribution and MI engine for mixed data.    
One approach is to assume the variables arise as in \eqref{eq:ymodel} and \eqref{eq:dx} with shared components 
$\Hxi=\Hyi\equiv H_i$, that is, 
\begin{gather}
(Y_i\mid H_i=h,-) \sim N(\mu_h, \Sigma_h),\ 
\Pr(X_i = x_i \mid H_{i} = h, -) =\prod_{j=1}^{p}\psi_{hx_{ij}}^{(j)}.
\end{gather}
This model makes strong \emph{local independence} assumptions, namely that $Y \id X \mid H$. This puts a significant burden on 
the mixture components. They must simultaneously capture non-normality in the distribution of $Y$, dependence between
 $Y$ and $X$, and dependence within $X$.  Doing so typically requires a large number of components and a commensurate 
amount of data. For example, since $X$ is categorical the true mean function can be written as ${E(Y\mid X=x)=\tilde D(x)\tilde{B}}$, where $\tilde D(x)$ is the true design vector and $\tilde{B}$ is the matrix of true regression coefficients.
The 
model has to include components at each distinct value of $\tilde D(x)\tilde{B}$---for all possible values of $x$---just to model the mean function, with further 
components to capture non-Gaussian structure in $Y$ and dependence in $X$. 

This burden can be alleviated somewhat by allowing the means to depend on $X$ as in the general location model: $\mu_h(x) = D(x)B_h$, with $D$ encoding main effects and possibly interactions. However, when $p>q$ as is common in survey data, the number of components required to adequately model dependence in 
$X$ tends to be much larger than that required to model $Y$, particularly since this model allows for local dependence 
in $Y$ (through the covariance matrices) but not in $X$.

An alternative approach is to use separate component indices and independent 
prior distributions for $\Pr(\Hxi=\hx)$ and $\Pr(\Hyi=\hy)$, as in \eqref{eq:ymodel}-\eqref{eq:Hip}, but add $(X,Y)$ dependence by setting $\mu_\hy(x) = D(x)B_{\hy}$.
We have 
 \begin{gather}
(Y_i\mid X_i = x_i, \Hyi=\hy,-) \sim N(D(x_i)B_\hy, \Sigma_\hy),\label{eq:ymodelwithx}\\
 \Pr(X_i = x_i \mid \Hxi = \hx, -) =\prod_{j=1}^{p}\psi_{\hx x_{ij}}^{(j)}.\label{eq}
 \end{gather}
This model enforces restrictive assumptions about the relationship between $Y$ and $X$. For example, we would 
have $E(Y\mid X=x) =  D(x)\left[\sum_{\hy=1}^{\ky} B_\hy\byy{\phi_\hy} \right]$, so that the model is unable to capture interactions not 
already coded in $D(x)$.

To construct the \ac, we use \eqref{eq:ymodelwithx} and \eqref{eq} as the data models.  However, rather
than choose between common components or independent components, we use 
a hierarchical prior distribution that maintains the desirable features of both, while 
incorporating new forms of local dependence. We now outline this hierarchical prior distribution for $(\Hxi, \Hyi)$. 

\subsubsection{Hierarchical prior for component indexes}\label{sec:indexmodel}

Let $Z_i$ be a third component index such that $1\leq Z_i\leq \kz$.  We assume $\Hxi$ and $\Hyi$ are independent given $Z_i$, so that
\begin{gather}
\Pr(\Hxi=\hx, \Hyi=\hy \mid Z_i=z) 
 = \byx{\phi_{z\hx}}\byy{\phi_{z\hy}}\label{eq:hmodel}\\
 \Pr(Z_i=z) = \lambda_z.\label{eq:zmodel}
\end{gather}
Here each $\byx{\phi_z} = \left(\byx{\phi_{z1}},\dots,\byx{\phi_{z\kx}}\right)'$ and
$\byy{\phi_z} = \left(\byy{\phi_{z1}}, \dots,\byy{\phi_{z\ky}}\right)'$ are probability vectors.
Both are assigned independent truncated stick breaking priors.  For $1\leq z\leq \kz$, we have
\begin{gather}
\byx{\phi_{z\hx}} = \byx{\xi_{z\hx}}\prod_{l<\hx}(1-\byx{\xi_{z\hx}}),\ \{\byx{\xi_{z\hx}}: 1\leq \hx\leq \kx-1 \}\iid Beta(1, \byx{\beta}),\ \byx{\xi_\kx}\equiv 1.\\
\byy{\phi_{z\hy}} = \byy{\xi_{z\hy}}\prod_{l<\hy}(1-\byy{\xi_{z\hy}}),\ \{\byy{\xi_{z\hy}}: 1\leq \hy\leq \ky-1 \}\iid Beta(1, \byy{\beta}),\ \byy{\xi_\ky}\equiv 1.
\end{gather}
Marginalizing over $Z$ gives $\Pr(\Hxi=\hx, \Hyi=\hy) = \sum_{z=1}^{\kz} \lambda_z\byx{\phi_{z\hx}}\byy{\phi_{z\hy}}$, inducing dependence between the latent component membership indicators.

The top-level mixture probabilities $\lambda=(\lambda_1,\lambda_2,\dots,\lambda_{\kz})'$ are also assigned a truncated stick breaking process:
\begin{align}
\lambda_z = \byz{\xi}_z\prod_{l<z}(1-\byz{\xi}_l),\ \{\byz{\xi}_z: 1\leq z\leq \kz-1 \}\iid Beta(1, \alpha),\ \byz{\xi}_{\kz}\equiv 1.
\end{align}
\cite{Banerjee2013} establish that as $(\kz,\kx,\ky)$ all approach $\infty$, this is a well-defined prior distribution, which they call an infinite tensor factorization (ITF) prior. 
We assign $\alpha, \byx{\beta}$, and $\byy{\beta}$ independent Gamma prior distributions with shape and rate parameters equal to 0.5. A convenient strategy for choosing the truncation levels $(\kz,\kx,\ky)$ is to pick moderate initial values, increasing them if the number of occupied components approaches its upper bound. Appropriate values will depend on the dataset; for all the models fit in this paper we take $\kz=15,\ \kx=90$, and  $\ky=60$, which we found to be conservative {upper bounds. Generally, the \ac\ is insensitive to specific choices of $(\kz, \kx, \ky)$ provided that they allow for unoccupied components. }

\subsubsection{Data model priors}

We next specify prior distributions for the parameters in \eqref{eq:ymodelwithx} and \eqref{eq}.  For each $\psi_\hx^{(j)}$, we use independent Dirichlet distributions,
\begin{equation}
 \psi^{(j)}_{\hx} \iid Dir(\gamma_{\hx 1}^{(j)}, \dots, \gamma_{\hx d_j}^{(j)}).
\end{equation}
Reasonable default choices for the hyperparameters include setting $\gamma_{\hx}^{(j)} = (1,\dots,1)$ or 
$\gamma_{\hx}^{(j)} = (1/d_j,\dots,1/d_j)$.  Both represent relatively vague information about the within-component probabilities. In practice, we find that posterior predictive distributions are usually insensitive to this choice, and we
use the latter going forward. 

For the parameters in each $Y$-component, we use hierarchical normal-inverse Wishart priors. The hierarchical priors are an alternative to more restrictive models, recognizing that many components will have a relatively small number of data points and that elements of $B_{\hy}$ in particular may be poorly estimated. We have 
\begin{equation}
 \{(B_{\hy}, \Sigma_{\hy})\}\iid MatN( B_0, I, T_B)\times IW(v, \Sigma)
\end{equation}
\begin{equation}
 (B_0, \Sigma)\sim MatN( 0, I, \sigma^2_{0\beta}I)\times W(w, \Sigma_0).
\end{equation}
Here, $MatN(M, \Phi, \Sigma)$ is the matrix normal distribution, i.e. the distribution of the $p^*\times q$ dimensional matrix $M + \Phi^{1/2}\Omega\Sigma^{1/2}$ when $\Omega$ is $p^*\times q$ with $\omega_{ij}\iid N(0,1)$. We assume that $T_B =\diag(1/\tau_{1},\dots, 1/\tau_{q})$, and assign $\tau_1,\dots \tau_q$ independent $G(0.5, 0.5)$ priors. In applications we find the posterior predictive distributions to be insensitive to this choice. 
To complete the hyperprior, we use the fact that $E(\Sigma_h) = \frac{v}{w-q-1}\Sigma_0$. 
We center and scale each element of $Y$ marginally
and take $v=q+1$, $w=q+2$, and $\Sigma_0 = \frac{1}{q+1}I$ throughout. In sufficiently large samples, inferences are insensitive to the choice of $\sigma^2_{0\beta}$; we use $10$.

\subsection{Properties of the \ac\ }\label{sec:modelprop}

\begin{figure}
 \centering
 \includegraphics[trim=0cm 4cm 0cm 4cm, width=.45\textwidth]{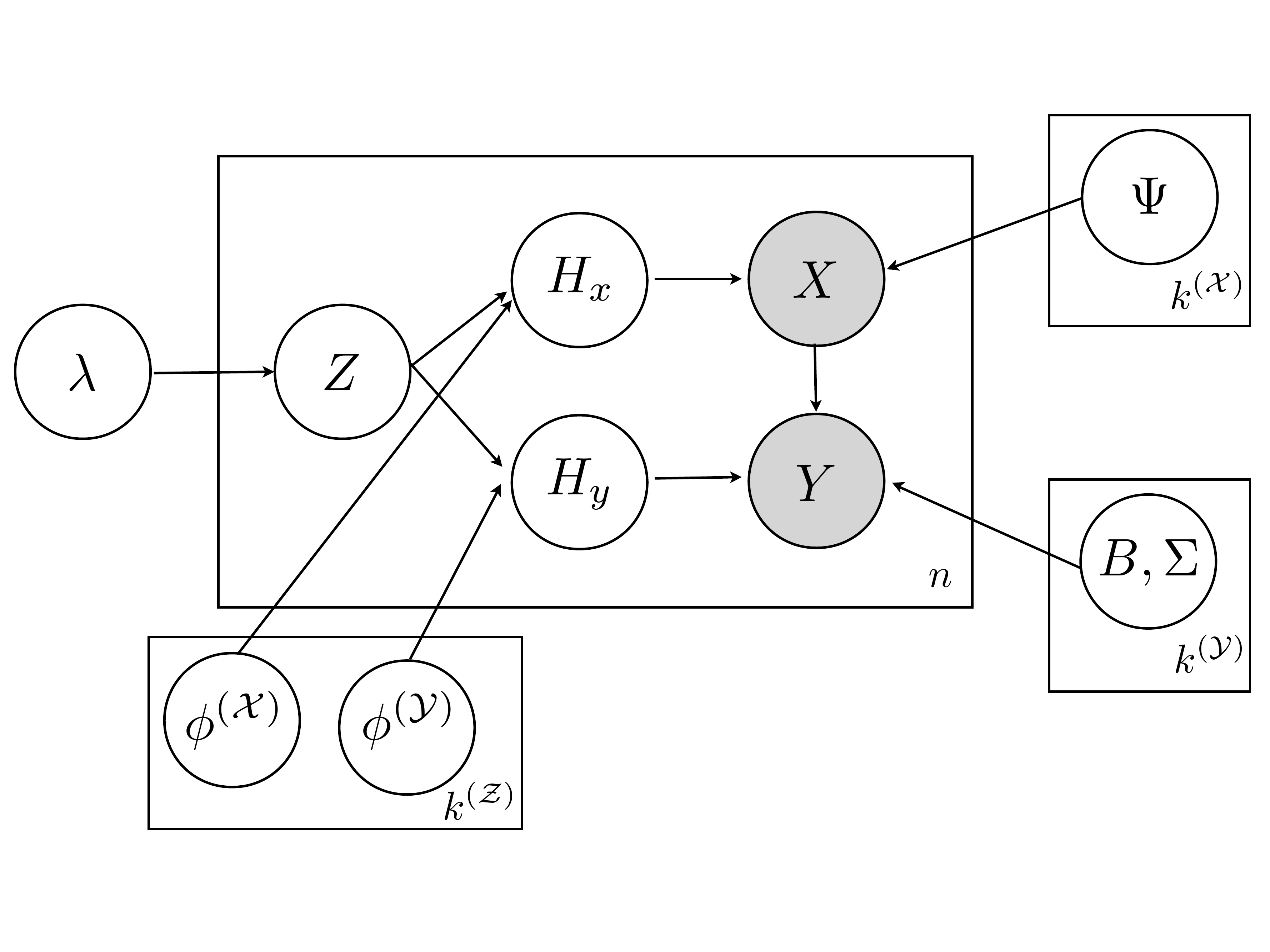}
 \caption{Model structure of the \ac.}
 \label{fig:plate}
\end{figure}

{Figure \ref{fig:plate} summarizes graphically the dependence structure of the \ac.} Marginally, $X$ has a latent class model {of the form
\begin{equation}
 \Pr(X_i = x_i) = \sum_{\hx=1}^{\kx} \left(\sum_{z=1}^{\kz} \lambda_z\byx{\phi_{z\hx}}\right) \prod_{j=1}^{p}\psi_{\hx x_{ij}}^{(j)},\label{eq:marx}
\end{equation}
where the term in parentheses gives the probability for class $s$}. This can capture any multivariate categorical distribution given sufficiently large $\kx$ (unlike unsaturated loglinear models). Conditional on $Y=y$, $X$ still follows a latent class model but with class probabilities that are functions of $y$. The  conditional distribution of $Y$ for any cell of the $X$ table is a mixture of multivariate normal distributions,
\begin{equation}
 f(Y_i\mid X_i=x_i) =  \sum_{\hy=1}^{\ky} 
 \frac{w_\hy(x_i)}{\sum_{l=1}^{\ky} w_l(x_i)} 
 N(Y_i; D(X_i)B_\hy, \Sigma_\hy)\label{eq:ygivenx}
\end{equation}
where $w_{\hy}(x_i) = \sum_{z=1}^{\kz} \lambda_z\byy{\phi_{z {\hy}}} 
\sum_{\hx=1}^{\kx}\byx{\phi_{z\hx}}\prod_{j=1}^{p}\psi_{\hx x_{ij}}^{(j)}$. 
The marginal distribution of $Y$ is also a mixture of multivariate normals.
 Thus, the \ac\ can represent a wide variety of shapes for the distribution of $Y$. 
Since $w_{\hy}(x_i)$ also appears in the expression for the conditional mean of $Y$,  
the \ac\ can
capture interactions not necessarily encoded in $D$.

The \ac\  {encodes} local dependence within and between $Y$ and $X$ in several ways{:} The $Y$-component specific regression 
functions and covariance matrices allow the relationships between $Y$ and $X$, and within $Y$, to vary by component. 
Further, the prior distribution in \eqref{eq:hmodel} - \eqref{eq:zmodel} implies that the \ac\ is a ``mixture of mixture models.''
Marginalizing over $\Hxi$ and $\Hyi$, the density of $(Y_i,X_i)$ given $Z_i=z$ is
\begin{equation}
f(X_i, Y_i \mid Z_i=z) = \left(\sum_{\hy=1}^\ky \byy{\phi_{z\hy}}N(Y_i; D(X_i)B_{\hy}, \Sigma_{\hy}) \right)\left(\sum_{\hx=1}^\kx \byx{\phi_{z\hx}}\prod_{j=1}^{p}\psi_{\hx X_{ij}}^{(j)}\right)\label{eq:xymidz},
\end{equation}
{so the joint density is
\begin{equation}
f(X_i, Y_i) = \sum_{z=1}^{\kz} \lambda_z\left(\sum_{\hy=1}^\ky \byy{\phi_{z\hy}}N(Y_i; D(X_i)B_{\hy}, \Sigma_{\hy}) \right)\left(\sum_{\hx=1}^\kx \byx{\phi_{z\hx}}\prod_{j=1}^{p}\psi_{\hx X_{ij}}^{(j)}\right)\label{eq:xy}.
\end{equation}
}

For any $z$, $(X\mid Z=z)$ follows an MPMN model.
 The distribution of $(Y\mid X, Z=z)$ is nearly the ANOVA-DDP model of \cite{DeIorio2004}, except that we relax 
their common covariance assumption with the hierarchical prior. From \eqref{eq:xymidz}, we see that within top-level components $Z$ 
we have local \emph{dependence} within $X$, as well as between $X$ and $Y$ (and within $Y$). Note that $f(X_i, Y_i\mid Z_i=z)$ and $f(X_i, Y_i\mid Z_i=z')$  differ 
only in their respective lower-level stick breaking weights, $(\phi^{(y)}_{z},\phi^{(x)}_{z})$ and $(\phi^{(y)}_{z'},\phi^{(x)}_{z'})$, and not in the lower-level parameters ($\{(B_\hy, \Sigma_\hy)\}$ and $\psi_\hx$). This is a parsimonious choice,  somewhat akin to assuming common covariance structures across components in normal mixture models (but much more flexible).

\subsection{Related work}\label{sec:mixlit}

\cite{Dunson2010} extended Dunson's and Xing's (2009) MPMN to mixed data by assuming fully factorized (product) kernels in a DP mixture. This 
model is a special case of the \ac\ with $\ky=\kx=1$, diagonal $\Sigma_\hy$, and $D(x_i)=1$. \cite{Dunson2010} note that when the number of variables grows the number of clusters also must grow to accommodate the dependence in the joint distribution. This is due to the local independence assumptions of the 
product kernel, which forces all the dependence to be represented through a single cluster index. As described in Section \ref{sec:modelprop}, the \ac\ is able to avoid such strong local independence assumptions through the use of multivariate normal regression components with full covariance matrices, as well as the structure imposed by the hierarchical prior on the mixture component memberships.

A number of authors have proposed joint mixture models that 
include limited local dependence to induce a
prior on one of the conditional distributions. These models decompose the joint kernel into a
conditional kernel for one variable given the others and a marginal product
kernel for predictors \citep{Shahbaba2009a,Molitor2010,Hannah2011}. 
Some of these are special cases of the \ac\ obtained by restricting $\ky=\kx=1$ and imposing more structure on the local covariances. These models are less relevant for imputation, where the entire joint distribution is of interest. Moreover, the assumption of local marginal independence between the majority of the variables can lead to the same proliferation of clusters as in 
latent class models \citep{Hannah2011}.

\cite{Banerjee2013} introduced the ITF prior to avoid the proliferation of components through \emph{dependent} component assignment in separate univariate mixtures. Compared to the \ac, this model encodes weaker local 
dependence, arising strictly through shared lower-level components within top-level components. 
\cite{Wade2011} proposed the enriched DP, which (like the ITF) models dependent cluster assignment. 
 The enriched DP separates a joint distribution into a conditional and a marginal, 
assigning each  a DP prior distribution where the base measure for the
conditional varies across the marginal. It lacks
the symmetry of the ITF, making it more difficult to interpret the
induced joint distribution and its margins. This is unappealing for our purposes.

The \ac\ has a number of benefits over existing alternatives. The hierarchical structure on component indices and other local dependence features allow the analyst to avoid the proliferation of clusters. At the same time, the induced marginal and conditional distributions are easy to derive and the \ac\ has appealing limiting forms (the MPMN model for $X$ and a multivariate regression or ANOVA-DDP for $Y\mid X$). Computation via MCMC is also straightforward, as detailed in the supplementary material. 

\section{Repeated Sampling Simulation Studies}\label{sec:sipp-sim}
To evaluate the performance of \ac\ in multiple imputation (MI), we
conducted several repeated sampling simulation studies on a constructed population
taken from the first wave of the 2008 SIPP panel. We define the 
population as individuals who reported positive income from work
during the reference period in Wave 1, excluding records with
missing entries. 
The constructed population consists of $N$=30,507
respondents.  With guidance from Census Bureau researchers, we select the two continuous
and eleven categorical variables displayed in
Table \ref{tab:sippvars}. We use a modest number of variables to
make a large repeated sampling study more efficient while keeping the problem
challenging. For example, the implied contingency table has over 7 million
cells and is very sparse.  %

Below we present comparisons of the \ac\ versus a fully conditional approach, as implemented in the R package \texttt{mice} \citep{VanBuuren2011}. We compare against a fully conditional approach as these have been repeatedly shown to perform at least as well as existing joint models for MI \citep[e.g.,][]{VanBuuren2007,lee2010multiple,kropko2014multiple} and are in widespread use. 
In the online supplement we also compare to a variant of the general location model; its performance is dominated by that of the \ac\ and MICE. The supplement also includes a simulation study comparing the \ac\ and MICE under missingness completely at random (MCAR). Overall conclusions under MCAR are similar to those below, but the performance difference is greater under MCAR since all variables were subject to missingnes in that case.

\begin{table}[t]
\caption{Variables in the repeated sampling simulation study}\label{tab:sippvars}

    \begin{tabular}{|l|l|}
    \hline
    Variable                           & Levels                                  \\ \hline
    Total monthly earnings from employment     & Continuous                              \\
    Age                                & Continuous                              \\
    Sex                             & 2                                       \\
    Race                               & 5                                       \\
    Marital Status                     & 6                                       \\
    Born in the US                     & 2                                       \\
    Number of own children in the home & 4 (0,1,2, or 3+)                        \\
    Education level                    & 6                                       \\
    Occupation                         & 23                                      \\
    Worker Class                       & 3 (Private, Nonprofit, Government)      \\
    Union                              & 2                                       \\
    Hourly                             & 2                                       \\
    Usual Hours worked                 & 9 (0-80 in increments of 10 hours, 80+) \\ \hline
    \end{tabular}
\end{table}

\subsection{MAR simulation study design}
We create 500 datasets by taking simple random samples of size
$n=6,000$.  In each dataset, we let a random sample of 180 observations be complete cases. Age and sex are completely observed. Let $R_{i,\text{var}}=1$ when the variable ``var'' is missing for observation $i$ and $0$ otherwise. Define $U_{i} = \ind{\text{sex}_i=\text{``male''}}$. We sample
 $R_{i,\text{earn}}$ and $R_{i,\text{child}}$ from Bernoulli distributions with probabilities derived from 
\begin{align*}
\logit[\Pr(R_{i,\text{earn}}=1)] &= -0.25 + 0.5U_{i} - \left(\frac{\text{age}_i-25-25U_{i}}{25}\right)^2\\
\logit[\Pr(R_{i,\text{child}}=1)] &= -1.5U_{i} - \left(\frac{\text{age}_i - 40 + 10U_{i}}{30 + 10U_{i}}\right)^2.
\end{align*}
We partition the remaining variables into two blocks: demographic variables (race, marital status, born in US) and variables directly related to employment (education, occupation, worker class, union, hourly, hours worked). For each variable $j$ in the demographic block, we sample each $R_{ij}$ from  Bernoulli distributions with probabilities derived from 
\begin{equation}
\logit[\Pr(R_{ij}=1)] = -1 + 0.7R_{i,\text{child}} + 1.25\kappa_{ij},
\end{equation}
where $\kappa_i$ is a $3$-dimensional vector drawn from a normal distribution with mean 0, unit variances and all correlations equal to 0.3. For each variable $j'$ in the employment block, we sample each case's $R_{ij'}$ from Bernoulli distributions with probabilities depending on  $R_{i,\text{earn}}$ instead of $R_{i,\text{child}}$:
\begin{equation}
\logit[\Pr(R_{ij'}=1)] = -1 + 0.7R_{i,\text{earn}} + 1.25\omega_{ij'},
\end{equation}
where $\omega_i$ is a $6$-dimensional vector drawn from a normal distribution with mean 0, unit variances and all correlations equal to 0.3. %

In this MAR design, approximately 1/3 of the entries are missing for each variable (except age and sex) and approximately 5\% of cases are complete. %
The resulting imputation problem is challenging, as highly correlated variables are more likely to be missing simultaneously. 
The MAR mechanism results in biased available case estimates; for example, regressing log earnings on $U_m$ gives estimates of the coefficient around 0.25 (SE 0.02) in the available cases, compared to a true value of about 0.36.

\subsubsection{Generating imputations}
Within each of the 500 simulated datasets we create $M=10$ multiple imputations
with the \ac.  We use the default prior distributions described
in Section \ref{sec:model}, after standardizing the continuous
variables, and include main effects for each categorical variable in $D(X)$.
We estimate the \ac\ for each dataset using 200,000 MCMC iterations from the
Gibbs sampler described in the supplemental material, discarding the
first $100,000$ iterations and keeping the imputations from every
$10,000^{th}$ iteration thereafter. This is very conservative;
examination of a handful of datasets suggests that these numbers
could be reduced by at least half without impacting the results. In
practice, of course, imputers should carefully examine MCMC diagnostics of
relevant identified parameters, such as marginal means, quantiles, and
variances or covariances in the completed datasets. We ran the
simulations in a heterogeneous cluster environment, so the run times
varied.  As a reference, a 2014 MacBook Pro can complete 10,000 iterations of
the MCMC sampler in about 20 minutes. Our implementation could be made much more efficient; we discuss scalability in Section \ref{sec:conclusion}.

With each incomplete dataset, we also implement multiple imputation via
chained equations using the R package \texttt{mice}. Our goal is to compare default
applications of the software to a default application of the \ac, so we did not alter any of the options 
to \texttt{mice} other than to set $M=10$. The default 
procedure imputes continuous variables via predictive mean matching
\citep{Little1988} and uses logistic regressions to impute discrete
variables. Each conditional model includes a main effect for every
other variable.  After imputing the data with both procedures, we obtain MI inferences for a number of estimands using the methods in 
\citet{Rubin1987}.  We compute completed-data estimates and standard errors using the
\texttt{survey} package in R \citep{Lumley2004},
incorporating a finite population correction.%

\subsubsection{Evaluation metrics}

{We evaluate the competing imputation methods based on the performance of the MI pooled estimate and associated confidence interval 
for a range of estimands. For each estimate, the ``true'' value is the corresponding quantity computed in the population of $N$ individuals.
Ideally, under repeated sampling and realizations of the nonresponse process, and across a range of estimands, the imputation methods yield 
completed datasets for which 1) pooled estimates are approximately unbiased for the corresponding population quantities, and 2) 
pooled confidence intervals with level $\alpha$ contain their true population values at least $(1-\alpha)\%$ of the time \citep{Rubin1996}. 
Thus, while the proposed imputation method is derived from a Bayesian model, the ultimate evaluations are frequentist;
see \cite{Rubin1987,Rubin1996} for justification of this perspective.} 

\subsection{Results}

We begin by examining the means of log monthly earnings by age
(discretized into 10 year intervals except for $<18$, $18-25$, and
$65+$), sex and presence of own children. 
We restrict to cells in the
table formed by the three categorical variables with expected counts
of at least 30. We work on the log scale rather than with untransformed incomes, as 
the skewness of the income distribution makes normal approximations
more likely to hold.%

\begin{figure}

{\centering \includegraphics[width=.4\linewidth]{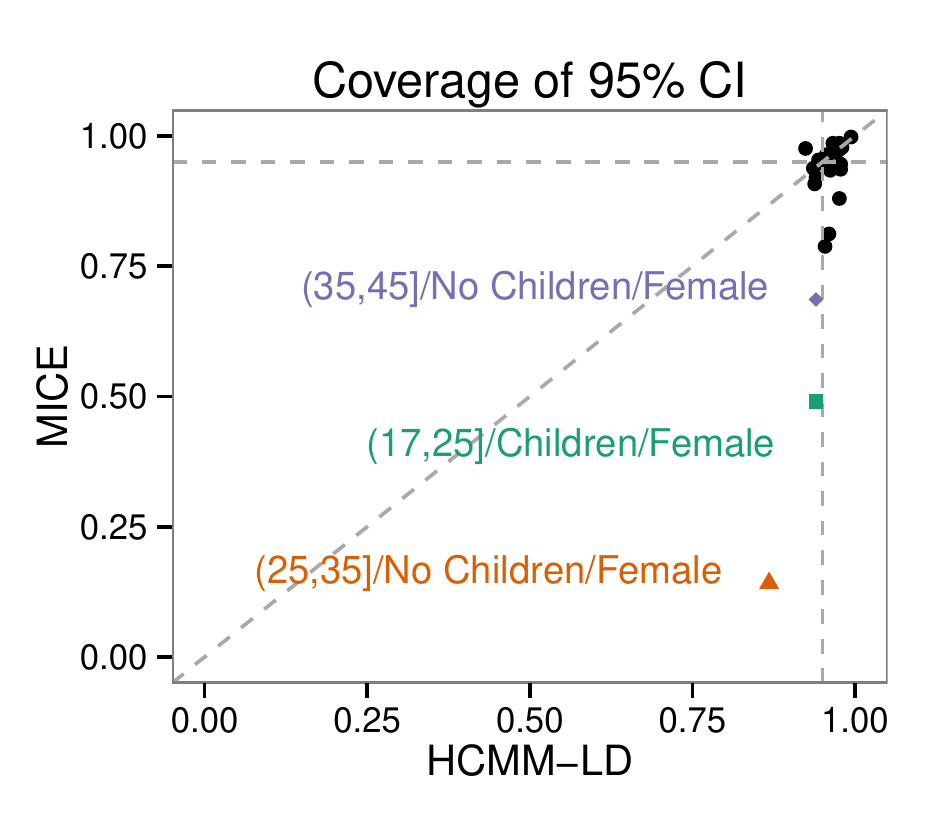} 
\includegraphics[width=.4\linewidth]{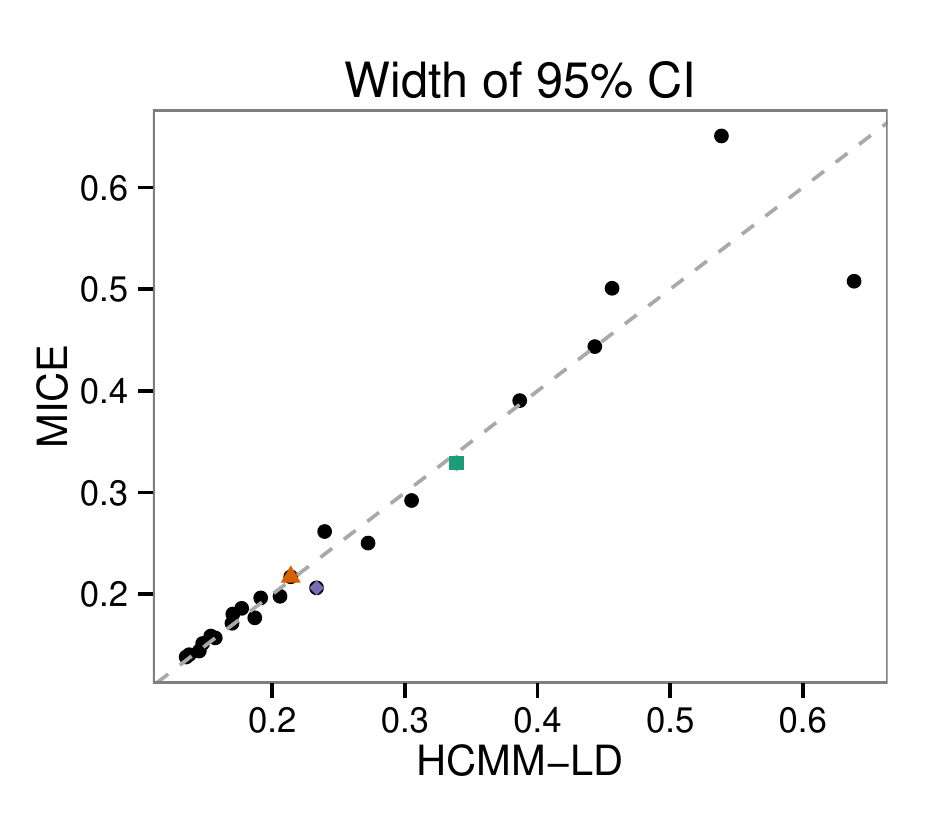} 

}
\caption{(Left) Coverage rate of pooled nominal 95\% CI for mean log monthly earnings by age, sex, and own children in the home (Yes/No) (Right) Average CI width
of 95\% CI.}
\label{fig:age-sex-kid-ci}
\end{figure}

\begin{figure}

{\centering \includegraphics[trim={0 0.5cm 0 0}, width=.35\linewidth]{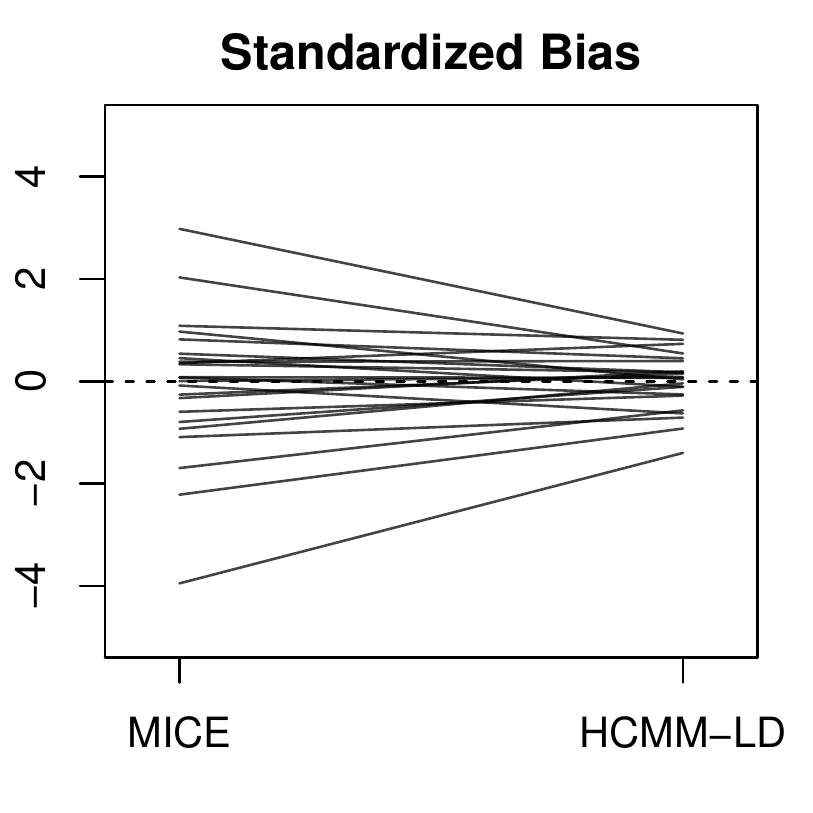} 
\includegraphics[trim={0 0.5cm 0 0}, width=.35\linewidth]{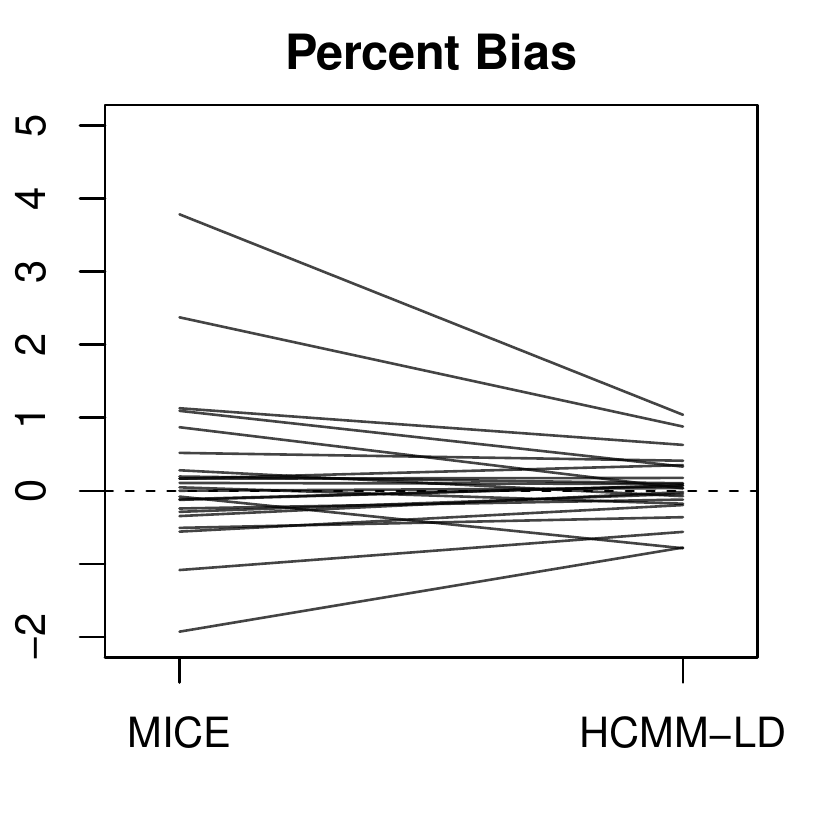} 

}

\caption{Standardized and percent bias of pooled
  estimates of mean log monthly earnings by age, sex, and own children
  in the home (Yes/No). Each line represents a cell mean, with the
  left and right endpoints at the bias under MICE and \ac,
  respectively.}
\label{fig:age-sex-kid-bias}
\end{figure}

Figure \ref{fig:age-sex-kid-ci} shows the coverage rates and average
width of 95\% multiple imputation confidence intervals. For most cells the pooled confidence intervals have approximately the correct coverage (the cluster of points around (0.95, 0.95) on the left hand side of Fig. \ref{fig:age-sex-kid-ci}). However, there are three cells where the MICE-constructed imputations yield substantially lower coverage than under the \ac. On average the pooled confidence intervals have similar widths, suggesting that bias in MICE's imputations drive the poor coverage. This is confirmed by Figure \ref{fig:age-sex-kid-bias}, which shows that while neither method has uniformly lower bias, the range of bias under the \ac\ is much smaller.

These are not especially small cells, although they do have somewhat higher probabilities of missingness for the own child and income variables (around 0.5 and 0.4, respectively). 
The same estimates are problematic in the MCAR simulation (see the supplementary material), so this is not merely a function of
 the MAR mechanism used here. 
Rather, it appears to be a function of complicated relationships among the variables involved.

Age and income have a relationship that the MICE imputations evidently
capture less effectively than the \ac\ imputations. For example, earnings tend to be lowest in the young (SIPP records earnings information on
respondents 15 or older), increasing during working years and falling
off again as those who can afford to retire do so. Additionally, the
variance in earnings is low in the younger cohort, roughly stable
through the working years, and increasing near and after retirement
age. Interactions also appear to be at play; the effect of
having their own child in the home varies across the respondent's age,
probably due in part to its high correlation with the age of the children, and
across the sexes as well. For example, the population difference in
log wages between those with children versus no children for 18-24 year old women is -0.159, whereas  for 35-44 year
old women it is -0.076. In men the population differences are -0.064 for 18-24 year olds and 0.232 for 35-44 year olds.

\subsubsection{Regression Coefficients}

\begin{figure}

{\centering \includegraphics[width=.4\linewidth]{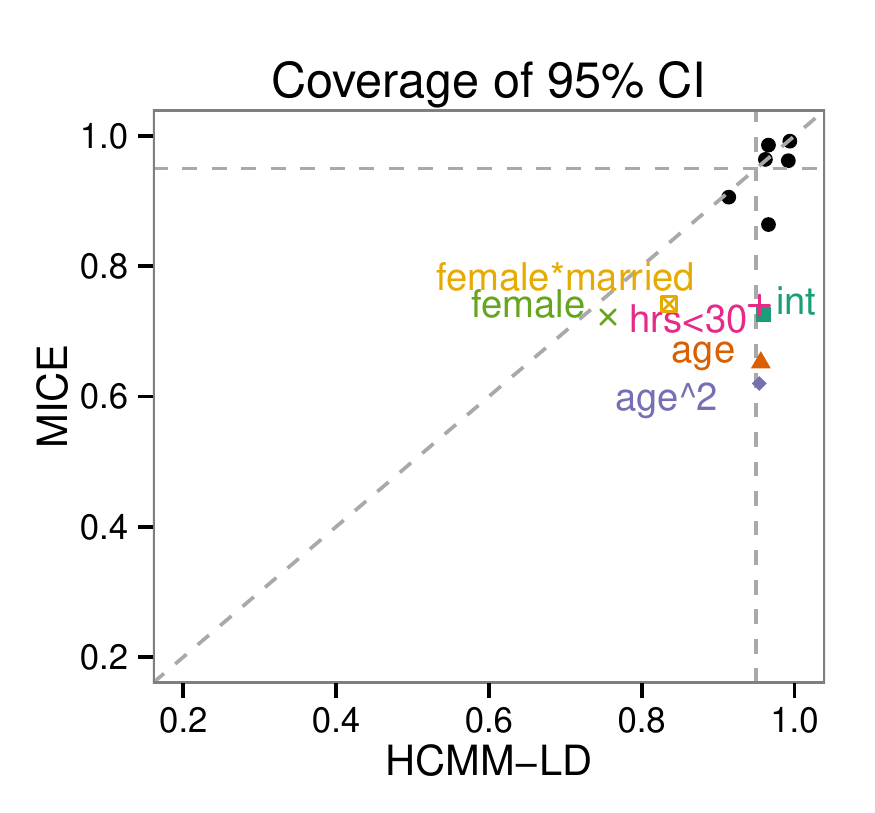} 
\includegraphics[width=.4\linewidth]{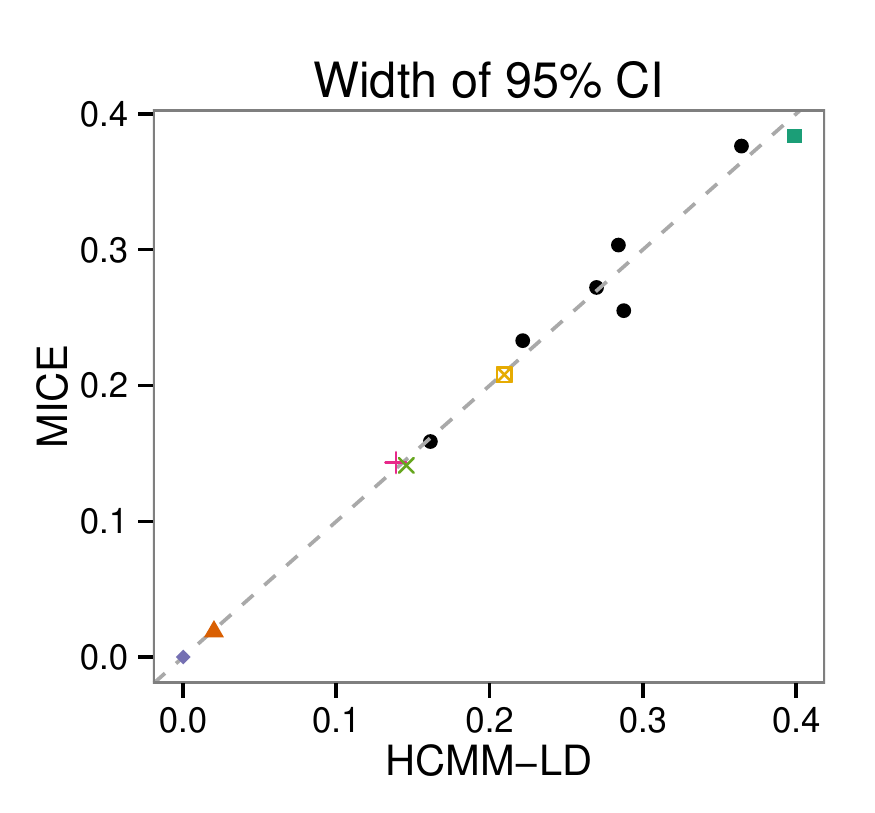} 

}

\caption{(Left) Coverage rate of pooled nominal 95\% CI for the regression with three-way interaction and age squared, including fpc. (Right) Average width
of 95\% CI.}
\label{fig:regms3sq}
\end{figure}

Next we consider linear regressions of log earnings on age, sex, usual
hours worked (recoded as $<30$, 30-60, and $60+$), and indicators for
marriage and own child under 18 in the
household. To begin we fit a model including an age squared term as 
well as two- and three-way interactions between sex, own child, and
marital status. Figure \ref{fig:regms3sq} displays MI
estimates of the coefficients and the average width of their
confidence intervals. {The \ac\ imputations result in better
repeated sampling properties overall. Neither method was modifed to anticipate the nonlinear relationship between age and income, although both have some ability to capture such relationships. MICE's predictive mean matching effectively borrows residuals from cases with similar predicted means, providing some protection against model misspecification \citep{Little1988}, whereas the \ac\ specifically intends to capture complex relationships between the variables.} From Figure \ref{fig:regms3sq} it is clear that the \ac\ is more successful at capturing this structure, with coverage nearer the nominal level for age and age squared (and all other coefficients), despite age being completely observed.

\begin{figure}

{\centering \includegraphics[width=.4\linewidth]{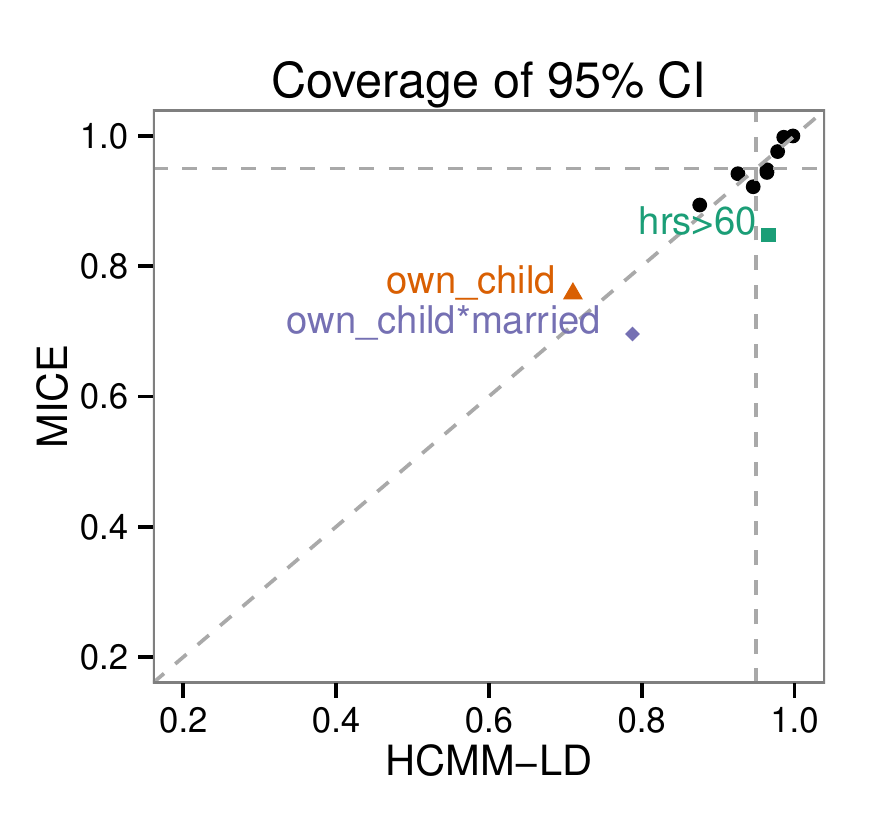} 
\includegraphics[width=.4\linewidth]{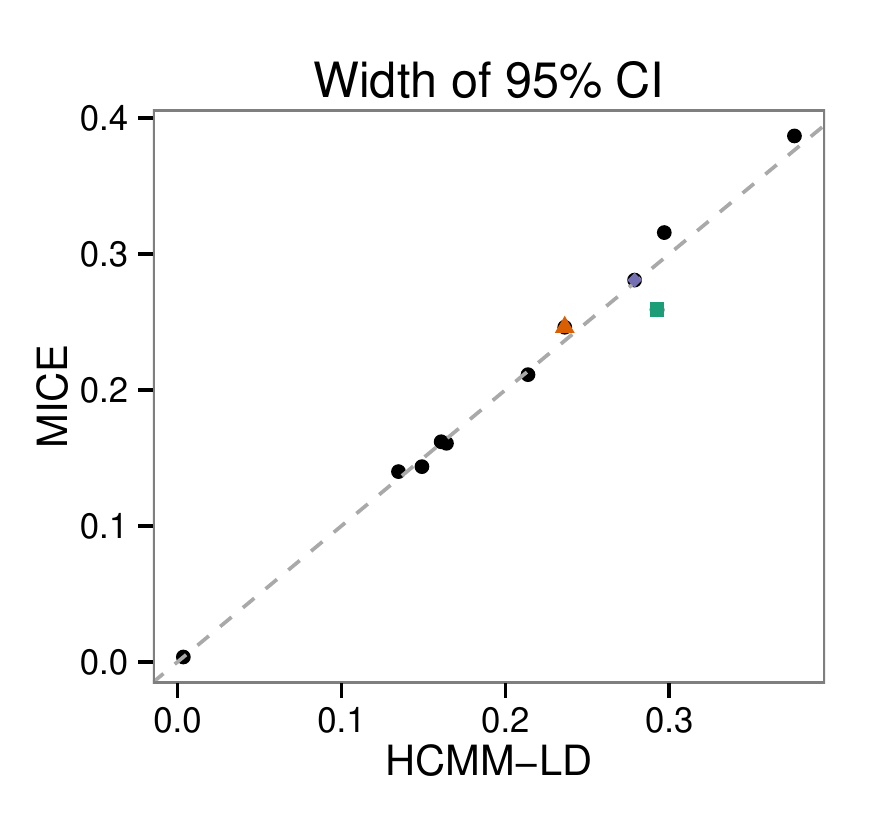} 

}

\caption{(Left) Coverage rate of pooled nominal 95\% CI for regression with three-way interaction \emph{without} age squared, including fpc. (Right) Average width
of 95\% CI.}
\label{fig:regms3}
\end{figure}

We also considered the same model excluding the age squared term.
Figure \ref{fig:regms3} shows that the results are largely similar for the former, with coverage generally improved overall but both methods struggling on the coefficients for own child and its interaction with the indicator of being married. Bias and average widths of confidence intervals are generally similar between the two methods on all the coefficients. A notable exception is the coefficient on the indicator variable for working over 60 hours, for which MICE achieves 85\% coverage to \ac's 96\%. The difference is driven by bias; the true population coefficient is 0.23, and the average pooled estimate using the MICE imputations is 0.18 versus 0.23 under \ac. The effect persists even in the model including only main effects (coverage of 82\% under MICE, versus 97\% under \ac). Here the true coefficient is 0.25, and the average point estimate from MICE is 0.18 versus 0.24 in the \ac. Coverage, widths of CIs and bias were essentially identical under both methods for the other coefficients. We had expected MICE to dominate in the main effects model since this is a submodel of the linear model MICE used to impute income. The relatively poor performance appears to
be due to the small sample size of this group (807 in the
population) and large true effect,
which combine to make predictive mean matching less effective.

\subsubsection{Conditional Frequencies}

\begin{figure}
{\centering \includegraphics[width=.4\linewidth]{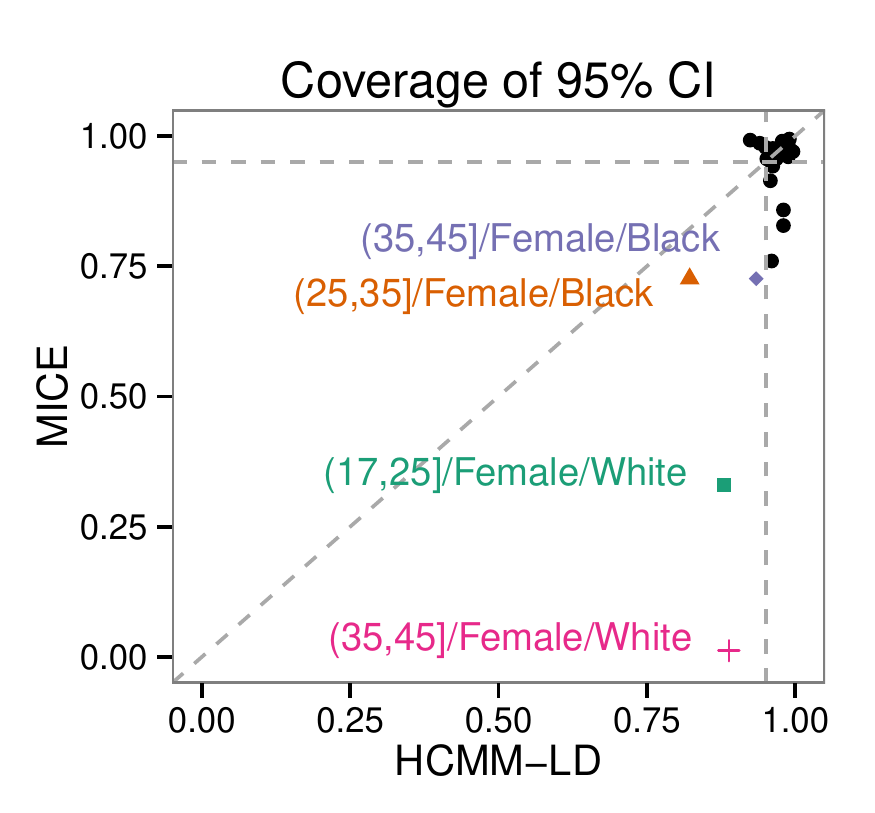} 
\includegraphics[width=.4\linewidth]{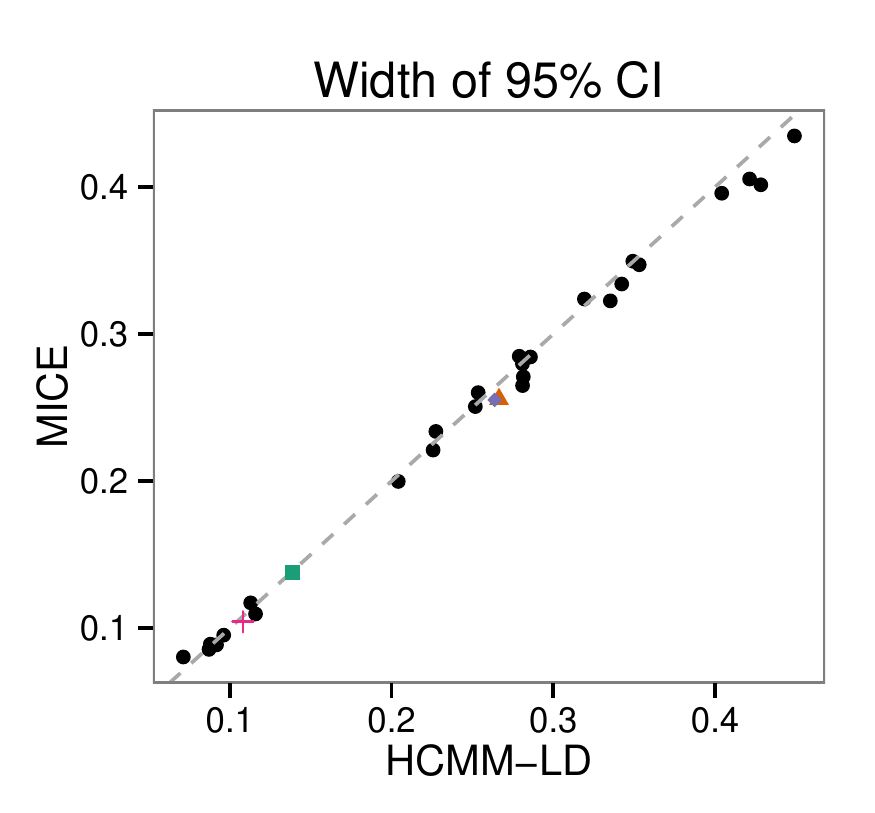} 

}
\caption{(Left) Coverage rate of nominal 95\% CIs for proportion with
  own child $<18$ in the household by age, race and sex. (Right) Average width 
of 95\% CI.}
\label{fig:prop_own_child}
\end{figure}

We also examine the quality of categorical imputations by estimating 
cell frequencies of categorical variables. We restrict to cases where
$E(n_c)\times p_c\geq10$ and $E(n_c)\times (1-p_c)\geq10$, where $p_c$ is
the true proportion and $n_c$ is the cell size in a simple random sample, to make the normal approximation more plausible. Figure \ref{fig:prop_own_child} displays results from estimating the proportion of respondents with their own child under 18 in the home by sex, race and age. Recall that only race and the presence of the respondent's own child have missing values.

\begin{figure}
{\centering \includegraphics[width=.55\linewidth]{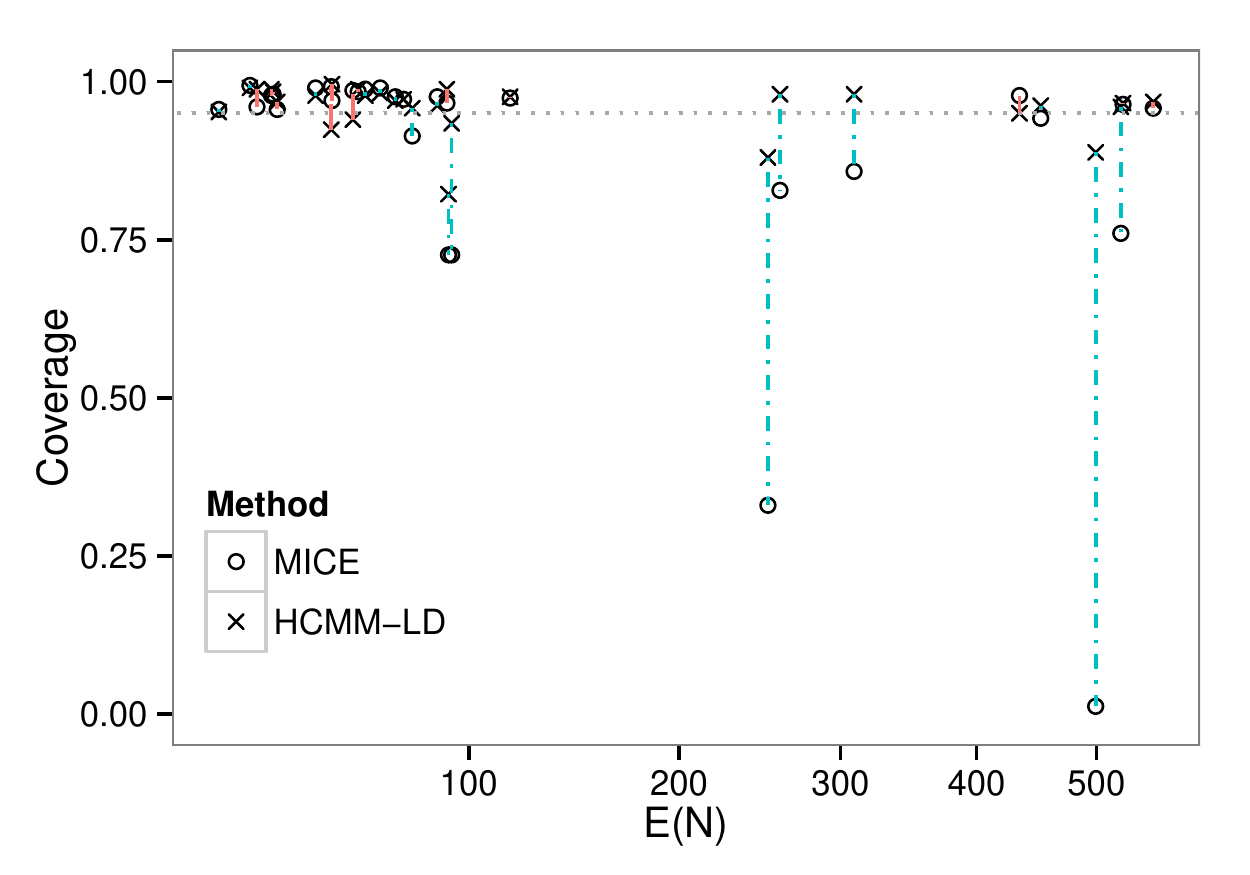} 

}
\caption{Coverage by expected cell size for proportion with own child $<18$ in the household by age, race and sex. Lines connect the coverage rates that correspond to the same estimand. Blue dot-dashed lines indicate that the \ac\ coverage is closer to 95\% than MICE, with red dashed lines indicating the reverse.}
\label{fig:prop_own_child_cov_by_n}
\end{figure}

The \ac\ based imputations perform much better than MICE. Coverage rates are uniformly better under the \ac\ with CIs of comparable width.
Coverage rates for the \ac\
never drop below 84\% (versus 71\% for MICE in that case), and the difference is dramatic for a few estimands. Figure \ref{fig:prop_own_child_cov_by_n} shows that MICE
has very good or very poor coverage in large cells, consistent with
the lack of coverage arising from misspecification bias. The \ac\
tends to have slightly lower coverage in these larger cells than in the smaller cells, but not nearly to the extent of MICE. 

\section{Evaluating the SIPP Redesign}\label{sec:msipp}

We now evaluate the agreement between the data from the field test and the data from the constructed sample of the
2008 production SIPP panel. For brevity, we use SIPP to refer to the subset of the production panel and SIPP-EHC to refer to 
the data from the field test. We focus on a subset of the data, namely household heads in 2010 from the SIPP-EHC and a 
contemporaneous wave of the SIPP subsample. The sample sizes are 2,588 for the SIPP-EHC and 3,665 for SIPP. 
Previously, Census Bureau researchers compared complete-case estimates from SIPP-EHC to those from production SIPP,
 and also to administrative records where available \citep{USCensus2013}. However, most 
 variables have missing values, and for some variables the missingness is substantial as evident in Table \ref{tab:msippvar}.

\begin{table}[tdp]
\begin{center}
\caption{Variables used from field test and production panel, and their fractions of missing data. An * indicates that the missing data percentage is computed as a fraction of the units known to be in-universe; for example, the percentage reported for monthly earnings is the fraction of respondents who indicated employment during the reference period but did not report the earnings amount. These also correspond to the continuous variables; the remainder are discrete.}
\label{tab:msippvar}

\begin{tabular}{l|l|l|l}

Variable Type & Variable & SIPP-EHC & SIPP \\
\hline
Household characteristics & Proxy interview & 0.00\% & 0.00\% \\
 & State & 0.00\% & 0.00\% \\
 & Household composition & 0.00\% & 0.00\% \\
 & No. persons in family & 0.00\% & 0.00\% \\
 & No. children under 18 & 0.00\% & 0.00\% \\
 \hline
Householder characteristics & Sex & 0.00\% & 0.00\% \\
 & Race/Ethnicity & 0.08\% & 4.12\% \\
 & Born in U.S. & 0.12\% & 0.05\% \\
 & Nativity/Citizenship status & 0.12\% & 0.05\% \\
 & Marital status & 0.70\% & 3.30\% \\
 & Disabled & 5.18\% & 4.09\% \\
 \hline
Work/Education & Educational attainment & 1.55\% & 0.00\% \\
 & Enrolled in school & 0.00\% & 0.00\% \\
 & Employment status & 15.96\% & 16.75\% \\
 & Monthly earnings* & 22.70\% & 17.56\% \\
 \hline
Program participation & Health insurance (any) & 2.43\% & 2.62\% \\
 & OASDI & 4.17\% & 1.77\% \\
 & SNAP & 1.85\% & 1.64\% \\
 & SNAP benefit amount* & 58.87\% & 65.00\% \\
 & TANF & 0.62\% & 0.22\% \\
 & SSI & 3.28\% & 3.98\% \\
 & Unemployment insurance & 4.13\% & 0.71\% \\
 \hline
Assets & Own interest bearing account & 6.34\% & 2.21\% \\
 & Own stocks/mutual funds & 5.80\% & 2.29\% \\
 & Own retirement account & 6.38\% & 1.88\% \\
 & Tenure in residence & 0.85\% & 0.03\% \\
 & Own home value* & 28.37\% & 38.57\% \\

\end{tabular}

\end{center}

\end{table}%

\begin{figure}
\begin{center}
\includegraphics[trim={0 0.9cm 0 0}, width=0.7\textwidth]{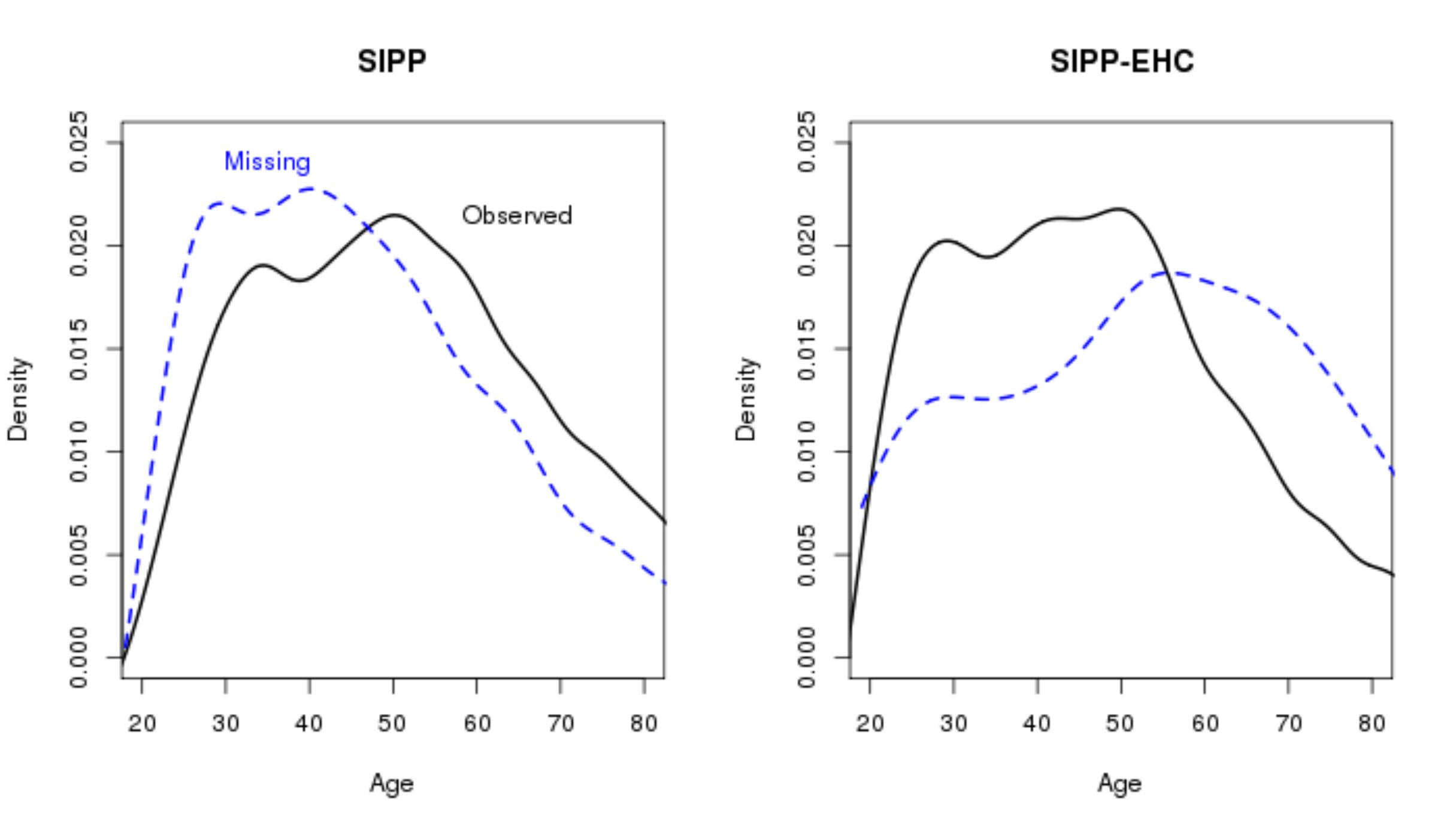}
\end{center}
\caption{Distributions of age in SIPP/SIPP-EHC by missing employment status indicator}
\label{fig:agemis}
\end{figure}

Missing data rates are generally similar in SIPP and SIPP-EHC, but missing data patterns vary substantially 
between the two surveys.  For example, 
Figure \ref{fig:agemis} shows density estimates of respondents' ages by whether their employment status is missing. 
In the SIPP sample, respondents with missing employment status are more likely to be younger, whereas in the SIPP-EHC 
they tend to be older. In each sample about 16\% of respondents are missing their employment status, so it is 
unlikely that these differences are due to sampling variability. In neither case is employment status plausibly missing completely at random, so comparisons based solely on complete cases or pairwise deletion 
are unreliable.

To account for missing data, we generated a set of $M=8$ completed datasets for SIPP and SIPP-EHC.
We used the HCMM-LD with the variables in Table \ref{tab:msippvar}, restricting the design vectors
to main effects only.
We imputed the SIPP-EHC and SIPP 
separately to avoid biasing the comparisons.
The continuous variables actually have a spike at zero corresponding to the 
unemployed, non-homeowners, or non-participants in SNAP. %
We decompose each of these into a binary indicator of a non-zero value and a continuous variable
 that is treated as missing anywhere the indicator is zero. Imputations for the original variable are constructed as the product 
of the indicator and the continuous variable (as in \cite{Heeringa2002}).

We run the MCMC for 130,000 iterations, discarding the first 50,000 and saving a completed dataset
 every 10,000$^{th}$ iteration thereafter. Standard MCMC diagnostics again suggest this is conservative. 
We compare the SIPP-EHC and SIPP on estimates of employment status, earnings, home value and SNAP benefit amounts,
 since these variables have significant fractions of missing data. We
 focus primarily on the effect that accounting for missing data has on comparing estimates from SIPP-EHC to the SIPP.

\subsection{Results: The impacts of accounting for missing data}\label{sec:SIPP-EHC:MI}

\begin{figure}
\begin{center}
\includegraphics[width=0.55\textwidth]{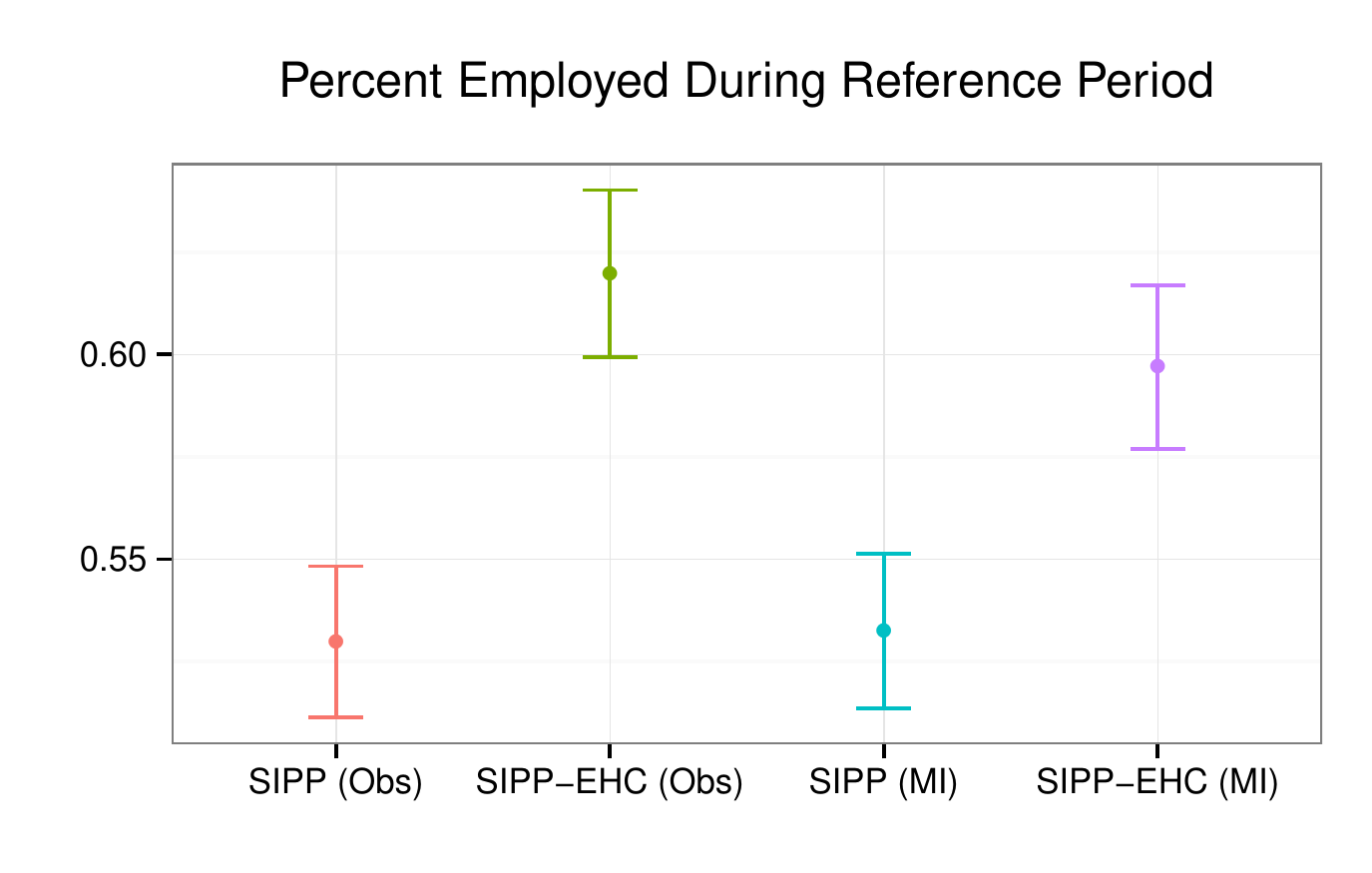}
\end{center}
\vspace{-20pt}
\caption{Percent reporting employment during the
  reference period, with 95\% confidence intervals, for SIPP-EHC and
the  production SIPP subsample.}
\label{fig:empstat}
\end{figure}

Figure  \ref{fig:empstat} displays estimates of the proportion of respondents who were 
employed at some point during the previous month, computed using the
\ac\ MI procedure and also using only the 
cases with observed values. Compared to the complete case estimates, the MI 
estimate is about 1\% higher for the SIPP and 1.5\% lower for
SIPP-EHC; hence, accounting for the missing data attenuates the
apparent differences in the estimates. 
This attenuation is concordant with the differential age distributions
of the cases with missing employment displayed in Figure \ref{fig:agemis}. Further, in SIPP-EHC individuals who 
report participating in SNAP are twice as likely to have missing income data as those who do not receive SNAP, whereas 
in SIPP the rate of missingness for employment status is about the same regardless of SNAP participation. Therefore, we 
expect more of those missing employment status in SIPP-EHC actually to be unemployed, and they are evidently imputed as such.

\begin{figure}
\begin{center}
\includegraphics[trim={0 0.5cm 0 0},width=0.95\textwidth]{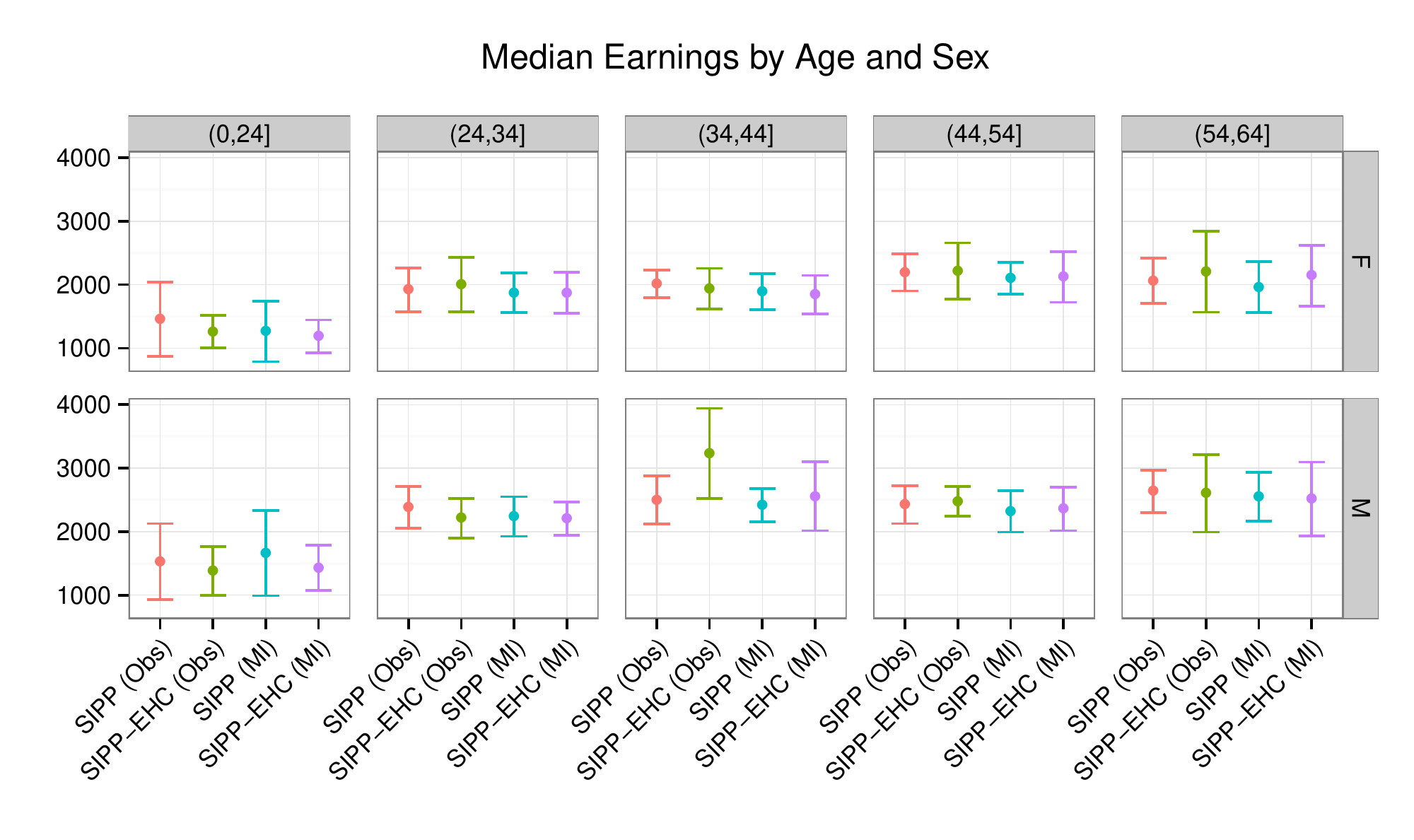}
\end{center}
\vspace{-20pt}
\caption{Median earnings by age and sex in SIPP/SIPP-EHC computed in complete cases and multiply imputed datasets}
\label{fig:medagesex}
\end{figure}

Figure \ref{fig:medagesex} displays similar quantities for the median
earnings by age and sex strata. For most strata, accounting for
missing data appears not to have substantial impact on the
comparisons (most estimates and 95\% CIs are similar pre- and
post-imputation). 
A notable exception occurs for 35-44 men: the MI estimate for SIPP-EHC is
substantially lower than the complete case estimate and appears more
in line with adjacent strata estimates.  
{In this particular cell, there are 122 respondents with observed values, whereas on average there are 172 men in the imputed datasets (these observations were missing either employment status or earnings amounts -- age and sex are completely observed). }%
Evidently, the \ac\ uses information from other covariates to generate imputations that are more in line with what one would expect, {pulling down the median in this cell to a value more consistent with neighboring estimates}.

\begin{figure}  

{\centering \includegraphics[trim={0 0.5cm 0 0},width=.45\linewidth]{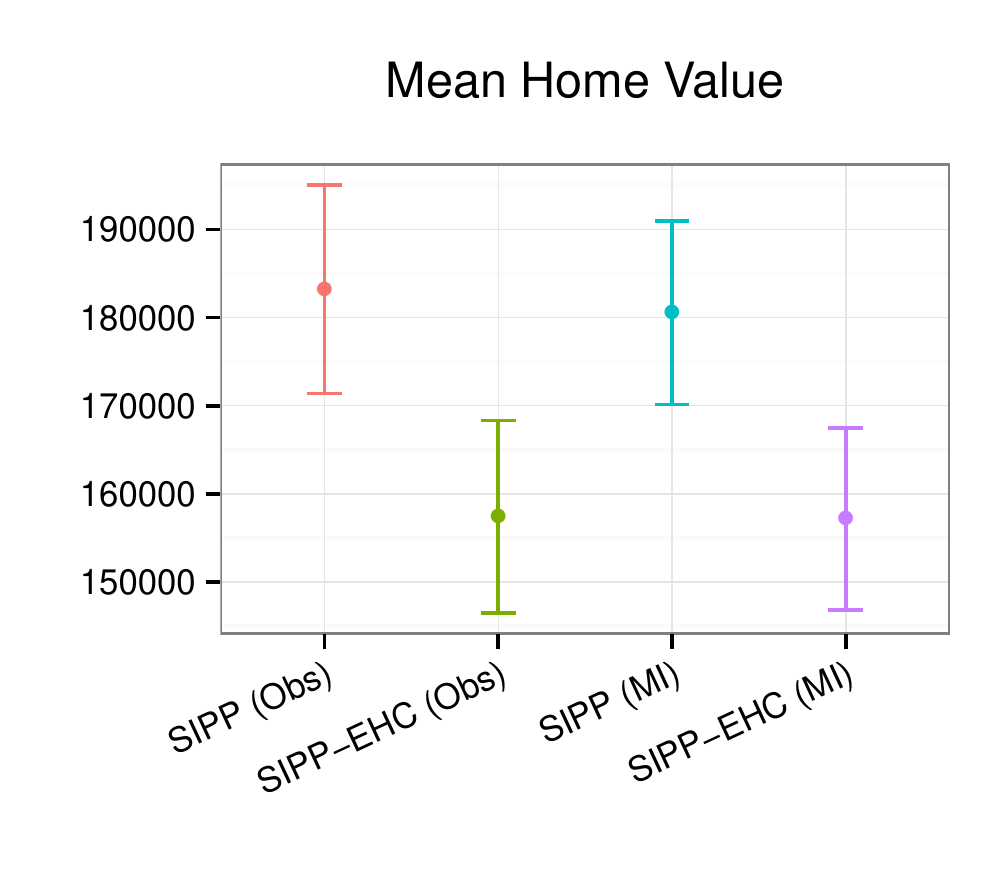} 
\includegraphics[trim={0 0.5cm 0 0},width=.45\linewidth]{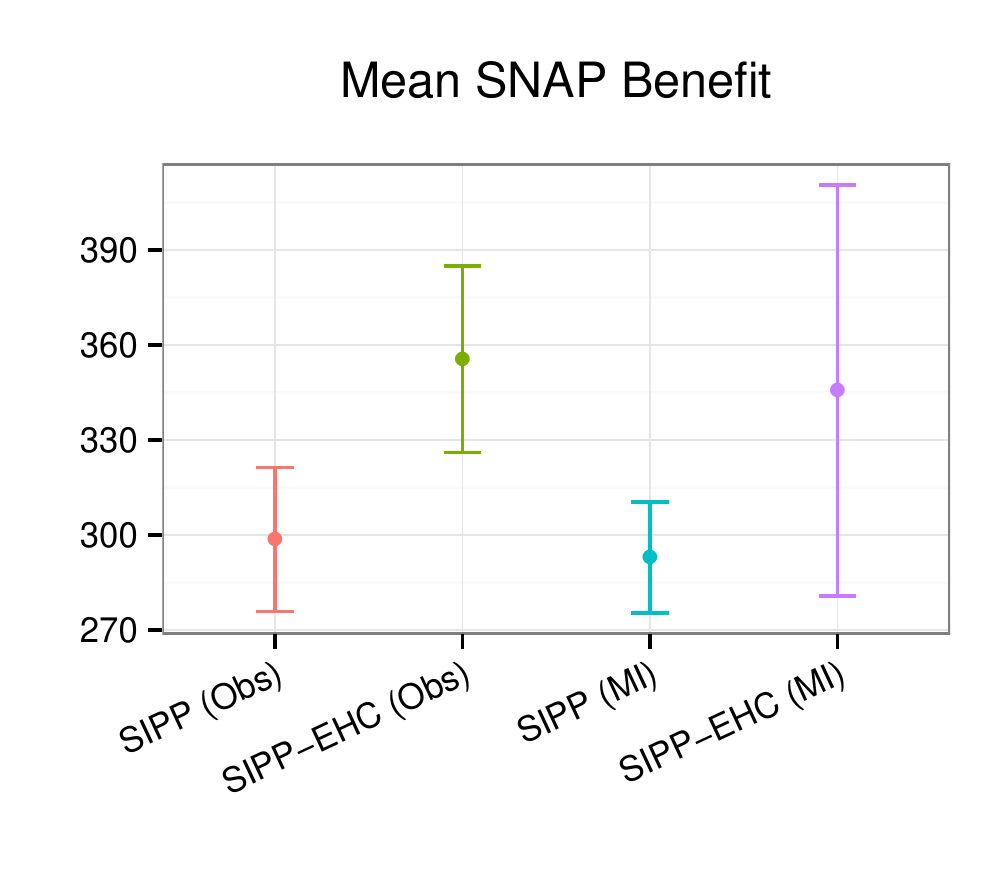} 

}  

\caption{Estimates of mean home value and mean SNAP benefit  in SIPP/SIPP-EHC computed in complete cases and multiply imputed datasets.}
\label{fig:snap-home}
\end{figure}

Finally, Figure \ref{fig:snap-home} shows imputed and complete case
estimates of mean home values and SNAP benefits. We focus on means
because these distributions are less skewed than the earnings
distributions. Imputed mean home values are very similar to complete
case estimates, with the differences much smaller than relevant
standard errors. 
The mean SNAP amounts are similar in the
imputed and corresponding complete case estimates, but for SIPP-EHC
the MI standard error is much higher than the complete case standard
error. This arises because of the relatively small sample
size; on average across imputations, 
there are about 560 respondents in universe for SNAP benefit amounts but only
225 have observed values, so about 60\% of the values are
missing. Moreover, about 25\% of SNAP
recipients are missing earnings data in SIPP-EHC, and earnings are one
of the most important determinants of SNAP benefit amounts (the other
is household size, which is completely observed). Although the rate of
missing amounts in the production SIPP subsample is about the same, SIPP has a lower
rate of missing income data among SNAP recipients (16\% versus
25\%) and a larger sample size (ranging from 893-901 across imputed
datasets). These factors combine to allow for  more precise
imputed estimates of SNAP benefits in the production SIPP subsample.

\subsection{Comparison of SIPP-EHC and production SIPP}

In their complete-case analyses, the Census Bureau researchers reported several
notable differences between the SIPP-EHC and production SIPP data \citep{USCensus2013}.  
For variables with fairly low rates of item nonresponse (5\% or
less), of course, accounting for item nonresponse with MAR models is
not likely to alter these conclusions.  
Our analyses of the variables with significant
missingness suggest nuanced conclusions about comparability.  In
particular, even after adjusting for item nonresponse, the SIPP-EHC
respondents are more likely to report being employed during the
previous month and also to report lower mean
home values.  Interestingly, Census Bureau researchers \citep{USCensus2013} linked complete cases
to administrative data and found that employment information was generally
more accurate in SIPP-EHC than in SIPP (these records
were not available to us).  Across the two samples, differences in
earnings and mean SNAP benefits tend to be small relative to MI
standard errors, particularly for SNAP benefits where relying on
complete cases appears to underestimate standard errors.

Despite being drawn from the same frame and weighted to the same population, the two data sources do exhibit some
differences among completely observed variables. For example, in
SIPP-EHC,  56.7\% of the householders are female (SD 0.9\%), compared to
60.3\% (SD 0.8\%) in the production SIPP subsample. In SIPP-EHC, the mean householder age is
47.8 (SD 0.34) compared to 50.4 (SD 0.29) in the production SIPP sample, and the quartiles
show a similar difference of about 2 years. A number of factors may
contribute to these differences; one candidate is differential
attrition or unit nonresponse between SIPP and SIPP-EHC, which seems
plausible given the substantial differences in their designs. The Census
Bureau is continuing to examine possible sources for this discrepancy
(personal communication). 

\section{Concluding Remarks}\label{sec:conclusion}

The repeated sampling simulation in  Section \ref{sec:sipp-sim} demonstrates that the \ac\ can serve
as a reliable default multiple imputation engine.  In fact, in the simulation
it often outperformed a default implementation of MICE, which is representative of the most widely used imputation routines. 
Of course, 
both MICE and the \ac\ could be modified to incorporate
dataset-specific prior knowledge -- and this is good practice -- but
in our experience many data users rely on default procedures. 
We examined many other potential estimands in the simulation study.
For many estimands the difference between default MICE and the \ac\
are modest, but for others the improvement under 
the \ac\  is substantial.  We suspect that the performance gap to increase as the
sample size grows, because the differences appear to be driven mostly
by misspecification bias. Unlike default MICE, as a nonparametric
Bayesian joint model  the \ac\ has the potential to increase in
complexity and capture additional features of the data. 

We have not performed a systematic evaluation of the
properties of the \ac\ in large sample, high-dimensional settings;
this is an area for future research.  We are optimistic about its
potential. Computationally, fitting the 
\ac\ reduces to fitting a series of mixture and regression
models. Computational complexity scales roughly linearly with sample size, dominated by computing likelihoods when resampling
cluster indices.%
These steps could be optimized further in our existing implementation. 
Increasing the dimension of the
categorical variables is clearly feasible; in an MPMN model \cite{Si2013}
considered simulations with 50 categorical
variables. Increasing the dimension of the continuous variables is more of a strain,
as the computation required grows quickly in
the dimension of the covariance matrices (for example, sampling the $\Hyi$ for $1\leq i\leq n$ is $O(n\ky q^2)$).

Fitting large
mixtures of multivariate normals is a well-studied problem, however, and specialized, efficient algorithms exist to leverage
parallel computing architectures (e.g. \citet{suchard2010understanding}). These would be
straightforward to adapt to the \ac. {So too are alternative parameterizations of the component-specific covariance matrices, 
such as factor-analytic forms in which $\Sigma_{\hy}= \Lambda_{\hy}\Lambda_{\hy}' + \Omega_\hy$ with $\Omega_\hy$ a diagonal $q\times q$ matrix and $\Lambda_{\hy}$ a $q\times b$ matrix of factor loadings (with $b<< q$). Such models reduce the number of free parameters and regularize the local covariances. They also render $Y_j$ and $Y_{j'}$ conditionally independent given some additional latent variables, simplifying imputation for  $Y$.}

There are a number of interesting directions to extend the \ac.  In
contexts with fully observed data, it can be advantageous to condition
on fully observed variables, such as design variables, so as not to
spend model fitting capacities on modeling these variables.  Since the
\ac\ is modular in nature, it is conceptually possible to incorporate
such adaptations.
Similarly, the modular nature suggests that it should be possible to
adapt the model to incorporate other types of variables, such as counts or durations. 
Finally, the current model does not account for structural zeros in contingency tables (from impossible
combinations or skip patterns), and linear restrictions
among continuous variables. We expect that it should be possible to
adapt the methods of  \cite{Manrique-Vallier2012,
  Manrique-Vallier2012a} to handle structural zeros and of
\cite{Kim2013} to handle linear constraints.  Adapting these approaches to mixed data
is an active area of research.

\appendix

\section{Supplementary Material}
In Section \ref{supp1}, we include results of an empirical study comparing the \ac\ to a variant of a general location model. In Section \ref{supp2}, 
we present results of a simulation study that uses a missing completely at random (MCAR) mechanism, comparing the \ac\ to MICE as in Section 4 of the paper. 
In Section \ref{supp3}, we describe the MCMC sampler for the \ac.

\section{Supplemental Simulation 1: General Location Model v.s. MICE under MAR}\label{supp1}

In the same missing at random simulation setting of Section 4 in the main text, we also attempted to implement a 
traditional general location model (GLOM) with loglinear constraints using the \texttt{mix} package in R \cite{Schafer1997}.  
Unfortunately, in this modest-sized problem \texttt{mix} ran into computational problems even when implementing reduced forms 
of the model (with common covariance, main effects for the linear regression and all two-way interactions for categorical variables). In the representation used by \texttt{mix}, the corresponding design matrix has over 7 million rows (one per cell), and the contingency table itself necessarily contains many sampling zeros.

We therefore considered a restricted version of the \ac\ that is similar to the traditional GLOM, obtained by setting $\kz=\ky=1$. In this case the model for
 $(Y\mid X)$ is a multivariate regression as in the traditional GLOM (with common covariance across cells), but the marginal distribution for $X$ is the MPMN model of \cite{Dunson2009} insteal of a loglinear model. This particular formulation has some advantages:
Compared to loglinear models for $P(X)$, the MPMN has the advantage of scaling to larger numbers of categorical variables and readily accommodating 
sparse tables \citep{Si2013}. 

We tested this restricted version of the \ac, which we will refer to simply as ``GLOM'' in this section, using the same MAR simulation setup as in Section 4 of the main text. We found that this model performed notably worse than the \ac\ and often worse than MICE as well. This is unsurprising, since the GLOM makes some assumptions that are questionable for these data, such as joint normality of age and logged earnings (conditional on $X$) (see also Section 2 of the main text).

Below we present results for several of the estimands we examined in Section 4 of the main text.  We display results for this GLOM relative to MICE,
with the intention of presenting evidence that MICE is a stronger competitor than GLOM for these data.

\subsection{Results}

As in Section 4 of the main text, we begin by examining the means of log monthly earnings by age
(discretized into 10 year intervals except for $<18$, $18-25$, and
$65+$), sex and presence of own children. 
We restrict to cells in the
table formed by the three categorical variables with expected counts
of at least 30. We work on the log scale rather than with untransformed incomes, as 
the skewness of the income distribution makes normal approximations
more likely to hold.%

\begin{figure}

{\centering \includegraphics[width=.4\linewidth]{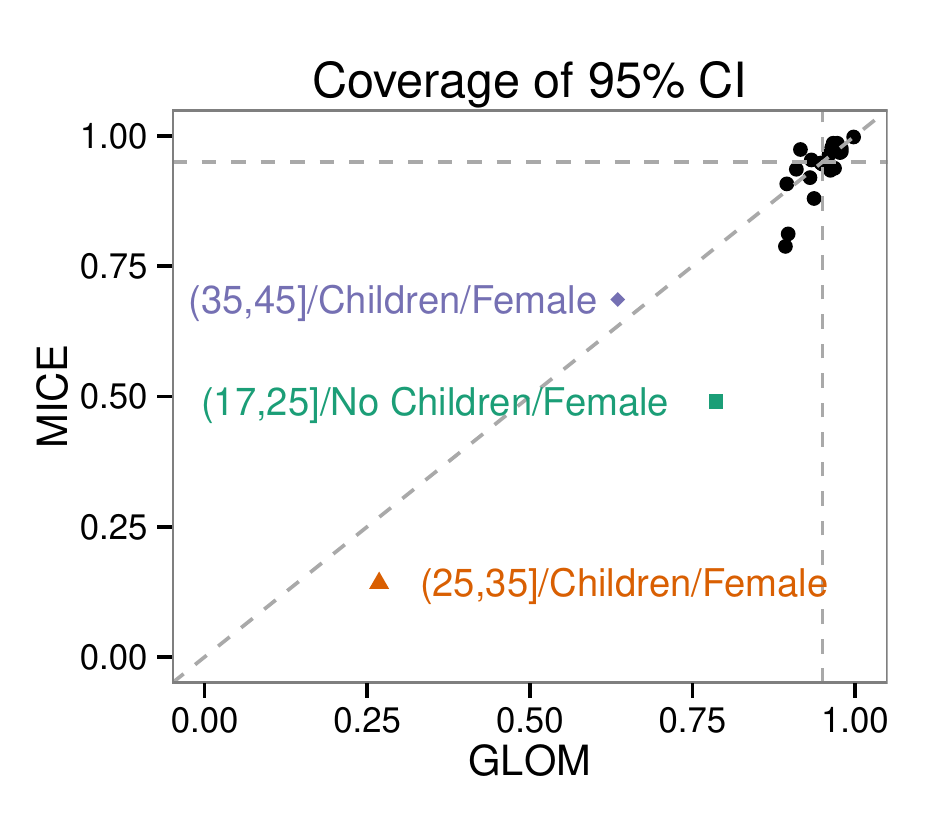} 
\includegraphics[width=.4\linewidth]{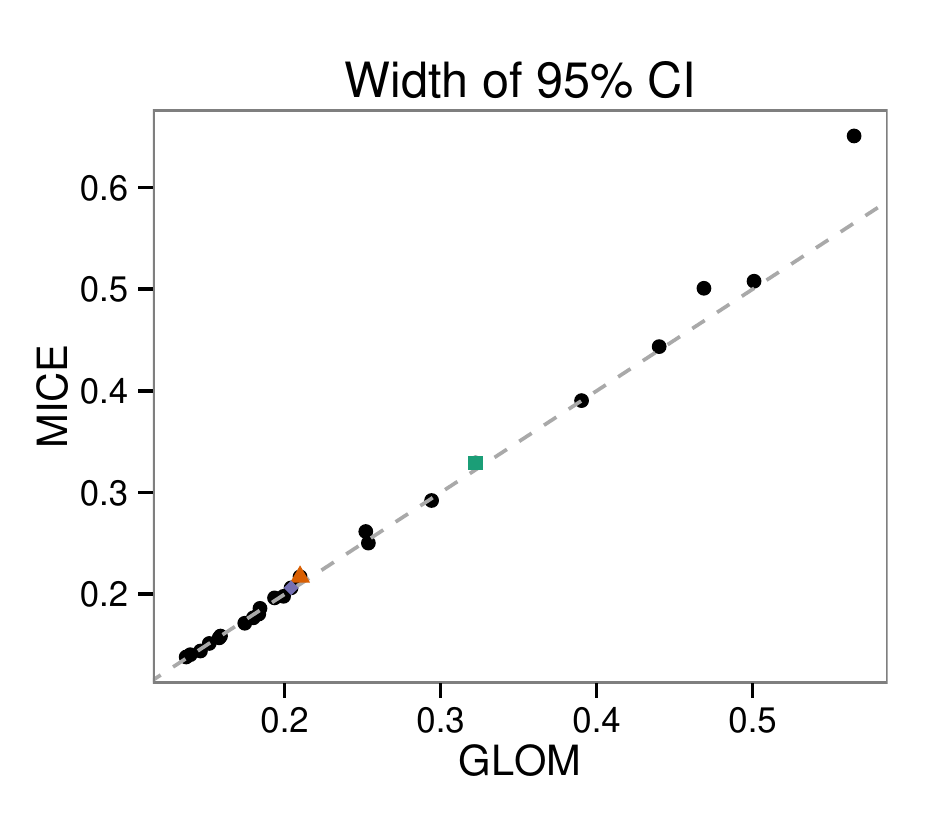} 

}
\caption{(Left) Coverage rate of pooled nominal 95\% CI for mean log monthly earnings by age, sex, and own children in the home (Yes/No) (Right) Average CI width
of 95\% CI.}
\label{sfig:age-sex-kid-ci-glom}
\end{figure}

Figure \ref{sfig:age-sex-kid-ci-glom} shows the coverage rates and average
width of 95\% multiple imputation confidence intervals. The GLOM generally outperforms MICE but not nearly as much as the full \ac, particularly on the three most troublesome cells (compare Figure \ref{sfig:age-sex-kid-ci-glom} to Figure 3 in the main paper). 

\subsubsection{Regression Coefficients}

\begin{figure}

{\centering \includegraphics[width=.4\linewidth]{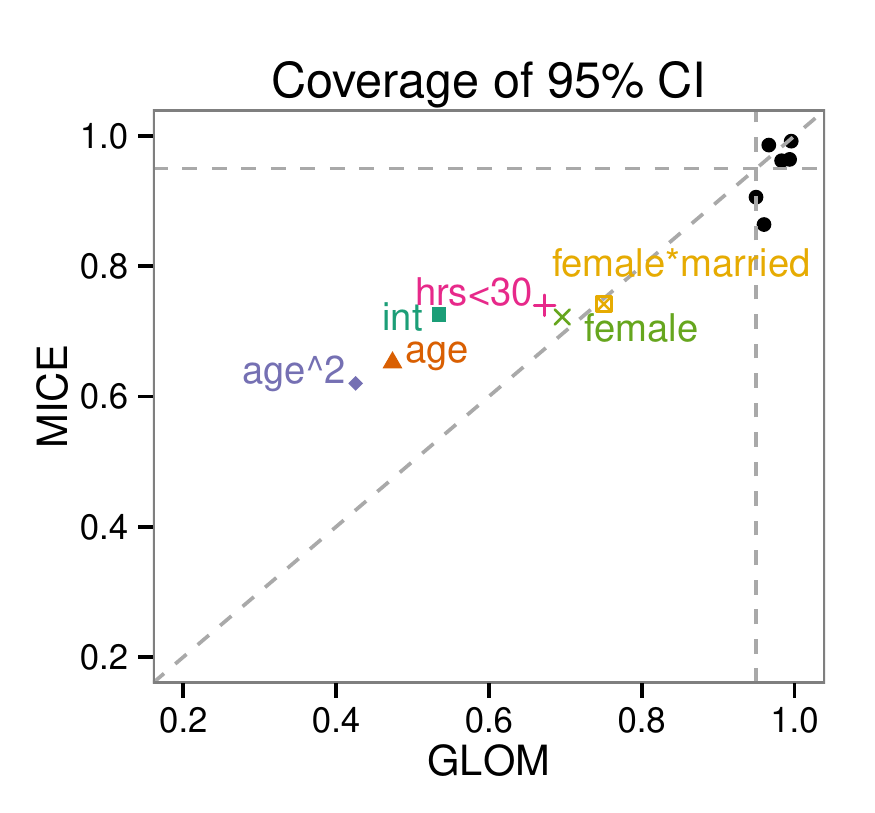} 
\includegraphics[width=.4\linewidth]{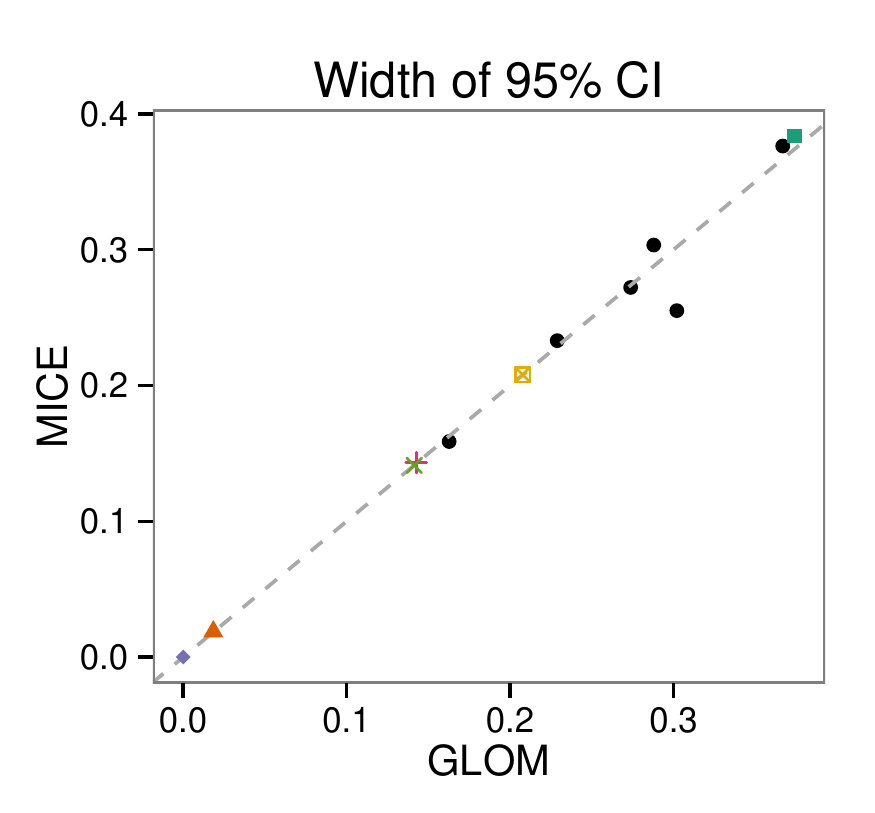} 

}

\caption{(Left) Coverage rate of pooled nominal 95\% CI for the regression with three-way interaction and age squared, including fpc. (Right) Average width
of 95\% CI.}
\label{sfig:regms3sq-glom}
\end{figure}

Next we consider linear regressions of log earnings on age, sex, usual
hours worked (recoded as $<30$, 30-60, and $60+$), and indicators for
marriage and own child under 18 in the
household. We fit a model including an age squared term as
well as two- and three-way interactions between sex, own child, and
marital status. Figure \ref{sfig:regms3sq-glom} displays MI
estimates of the coefficients and the average width of their
confidence intervals. 

Here we see a marked difference between the \ac\ and the GLOM. The \ac\ was able to accommodate the nonlinear relationship between age and earnings, attaining nominal coverage rates for the relevant coefficients (see Figure 5 in the main paper). The GLOM struggles here, performing worse than MICE. This should be expected, since unlike the \ac\ and MICE's predictive mean matching the GLOM has no capacity to capture nonlinear relationships in $Y$. On removing the squared terms, MICE, the \ac\ and the GLOM all perform similarly in the model with two and three way interactions. For the model with only main effects, both the GLOM and the \ac\ acheive nominal coverage rates for all the coefficients. As mentioned in Section 4.2.1 MICE does not, with coverage rates for the coefficient on the indicator of working over 60 hours a week dropping to 85\%.

\subsubsection{Conditional Frequencies}

\begin{figure}
{\centering \includegraphics[width=.4\linewidth]{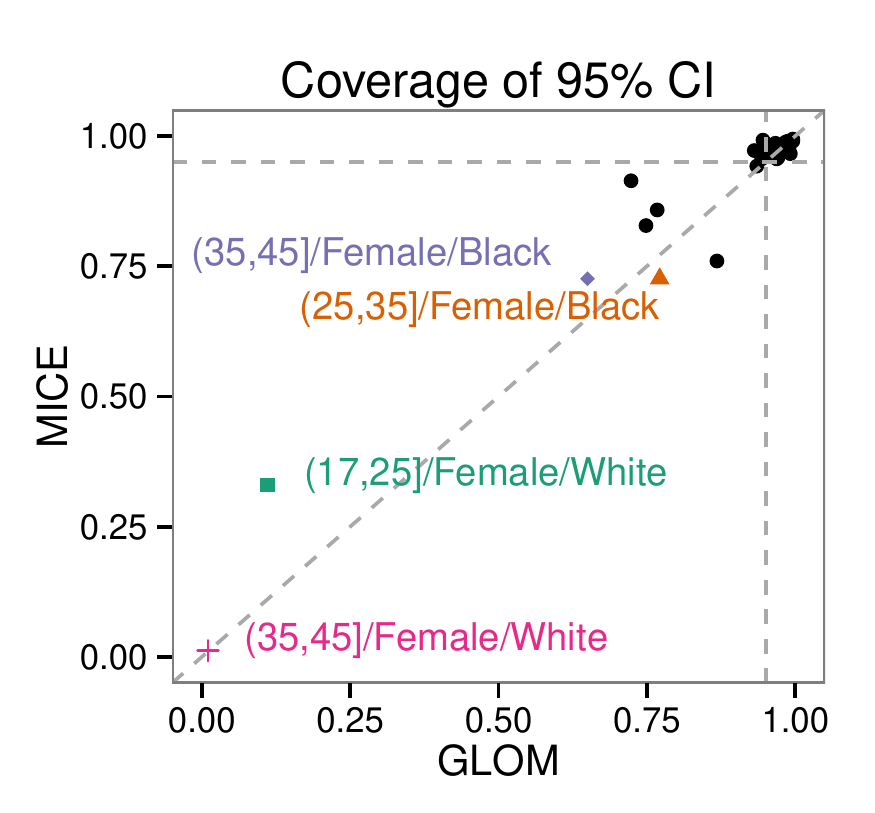} 
\includegraphics[width=.4\linewidth]{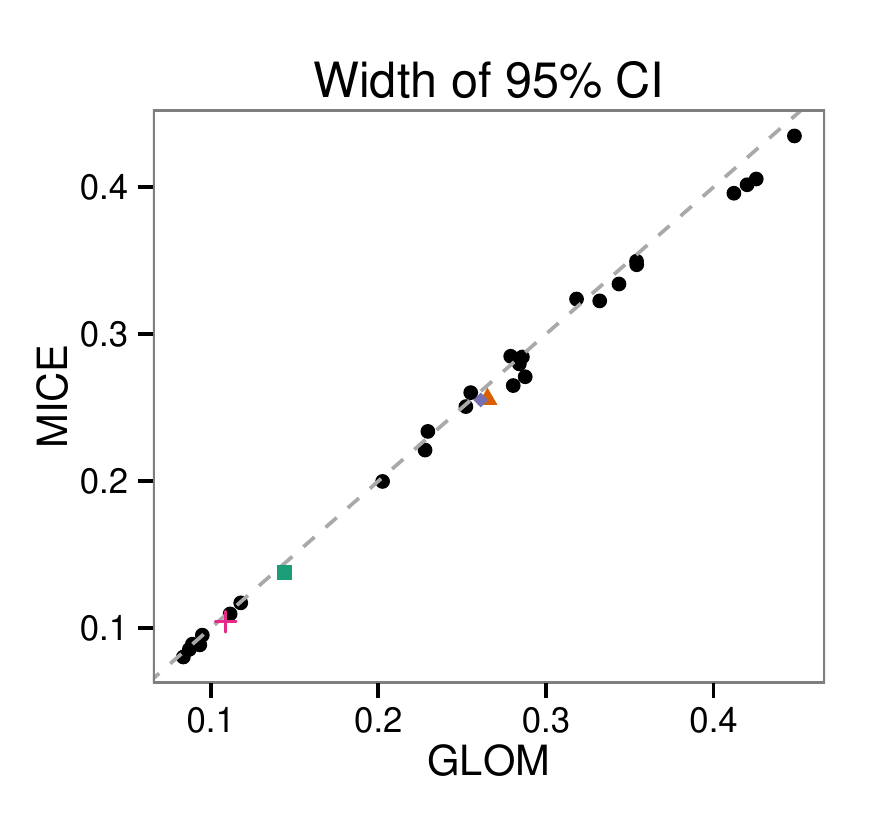} 

}
\caption{(Left) Coverage rate of nominal 95\% CIs for proportion with
  own child $<18$ in the household by age, race and sex. (Right) Average width
of 95\% CI.}
\label{sfig:prop_own_child-glom}
\end{figure}

We also examine the quality of categorical imputations by estimating
cell frequencies of categorical variables. We restrict to cases where
$E(n_c)\times p_c\geq10$ and $E(n_c)\times (1-p_c)\geq10$, where $p_c$ is
the true proportion and $n_c$ is the cell size in a simple random sample, to make the normal approximation more plausible. Figure \ref{sfig:prop_own_child-glom} displays results from estimating the proportion of respondents with their own child under 18 in the home by sex, race and age. Recall that only race and the indicator for presence of the respondent's own child are subject to missingness in this case.

\begin{figure}
{\centering \includegraphics[width=.65\linewidth]{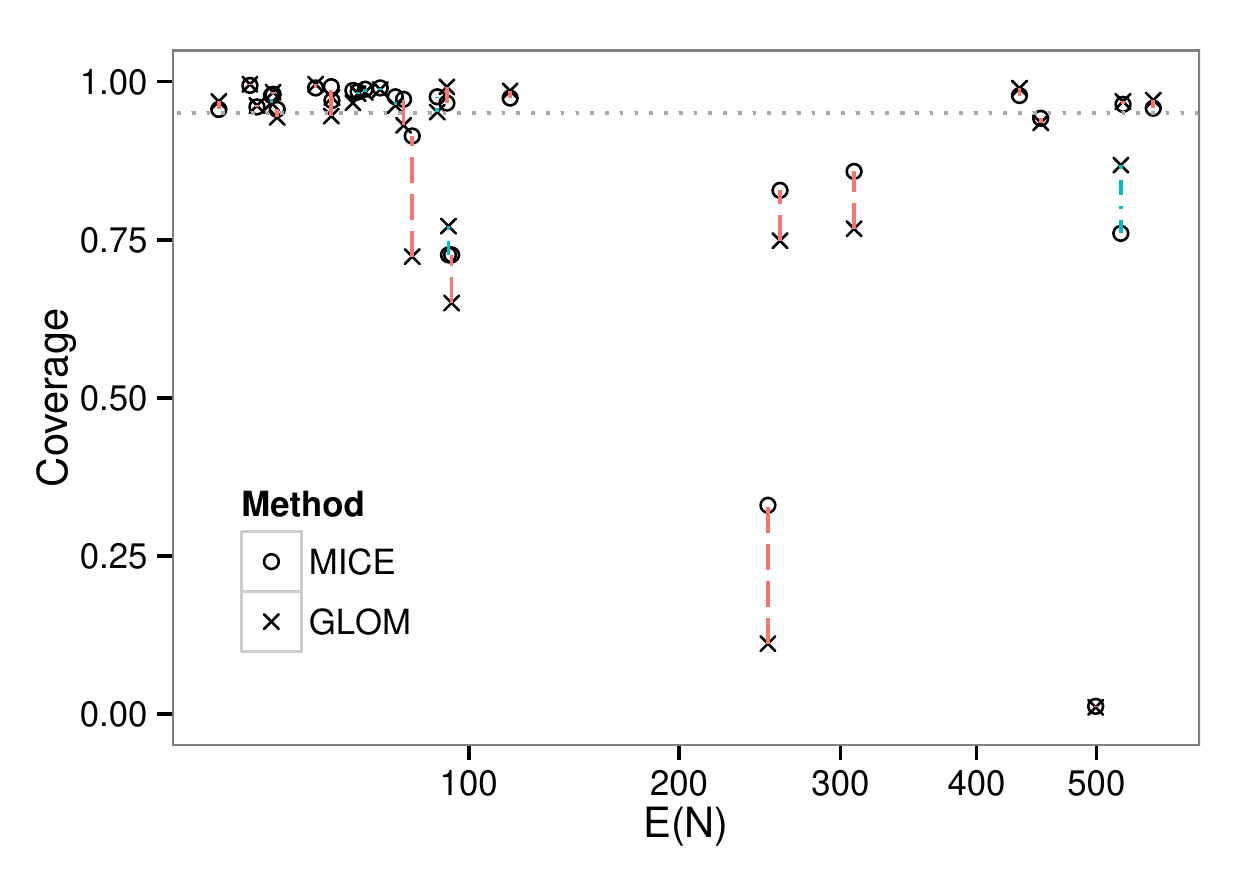} 

}
\caption{Coverage by expected cell size for proportion with own child $<18$ in the household by age, race and sex. Lines connect the coverage rates that correspond to the same estimand. Dot-dashed lines indicate that the \ac\ coverage is closer to 95\% than MICE.}
\label{sfig:prop_own_child_cov_by_n-glom}
\end{figure}

Overall, the GLOM and MICE imputations are similar. Again, this is in stark contrast to the imputations from the full \ac, which were uniformly better than MICE -- with coverage never dropping below 84\% -- and occasionally dramatically so (compare Figure 7 from the main paper to Figure \ref{sfig:prop_own_child-glom} here). Figure \ref{sfig:prop_own_child_cov_by_n-glom} shows that like MICE, the GLOM suffers from low coverage in the larger cells, which we may attribute to misspecification bias. In particular, we suspect that the rigid form of dependence between age and the categorical variables (only through the main effects of the categorical variables in the mean of age) is particularly problematic for the GLOM. 

\section{Supplemental Simulation 2: \ac\ v.s. MICE under MCAR}\label{supp2}

We repeat the simulation study in Section 4 of the main text but instead use a MCAR instead of MAR design. We use the same 500 simple random samples of size $n=6000$, but imposing missingness completely at random (with probability 0.35) after setting aside 500 complete cases. In this simulation all the variables (including age and sex) are subject to missingness. The rest of the simulation setup (truncation levels and MCMC iterations for the \ac, etc.) is the same as in Section 4.

Qualitatively, the results in the MCAR simulation are similar to those from the MAR simulation. 
The \ac\ and MICE are similar for many estimands, but the \ac\ is occasionally substantially better in terms of bias, coverage, and width of 95\% confidence intervals, and almost never worse.
The differences are more stark 
under MCAR, which we suspect is primarily due to the fact that age and sex are also subject to missingness in this study.

\subsection{Results}

We begin by examining the means of log monthly earnings by age
(discretized into 10 year intervals except for $<18$, $18-25$, and
$65+$), sex and presence of own children. We restrict to cells in the
table formed by the three categorical variables with expected counts
of at least 30. We work on the log scale rather than with untransformed incomes, as 
the skewness of the income distribution makes normal approximations
more likely to hold on the log scale. %

\begin{figure}

{\centering \includegraphics[width=.4\linewidth]{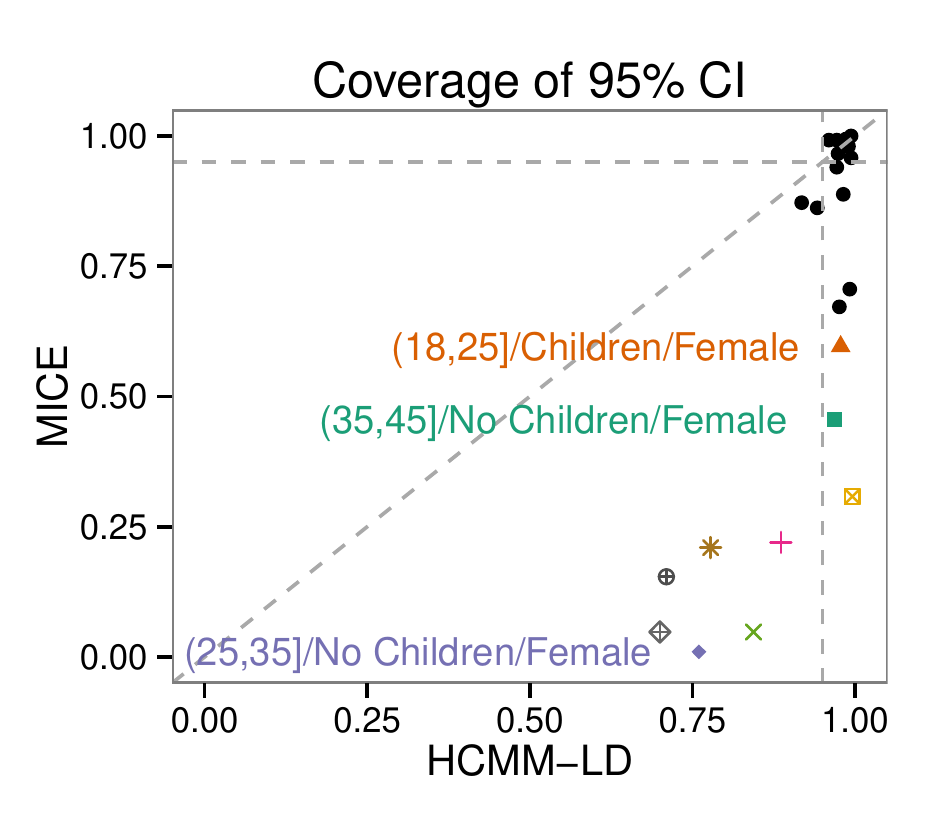} 
\includegraphics[width=.4\linewidth]{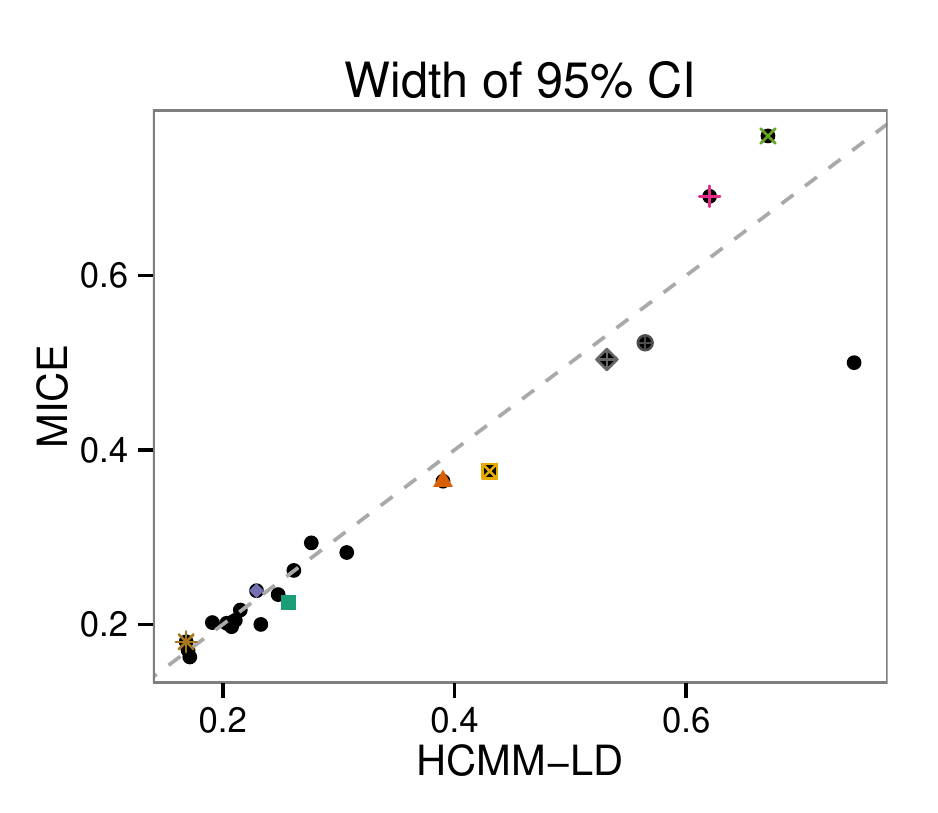} 

}
\caption{(Left) Coverage rate of pooled nominal 95\% CI for mean log monthly earnings by age, sex, and own children in the home (Yes/No) (Right) Average CI width
of 95\% CI.}
\label{sfig:age-sex-kid-ci}
\end{figure}

\begin{figure}

{\centering \includegraphics[width=.35\linewidth]{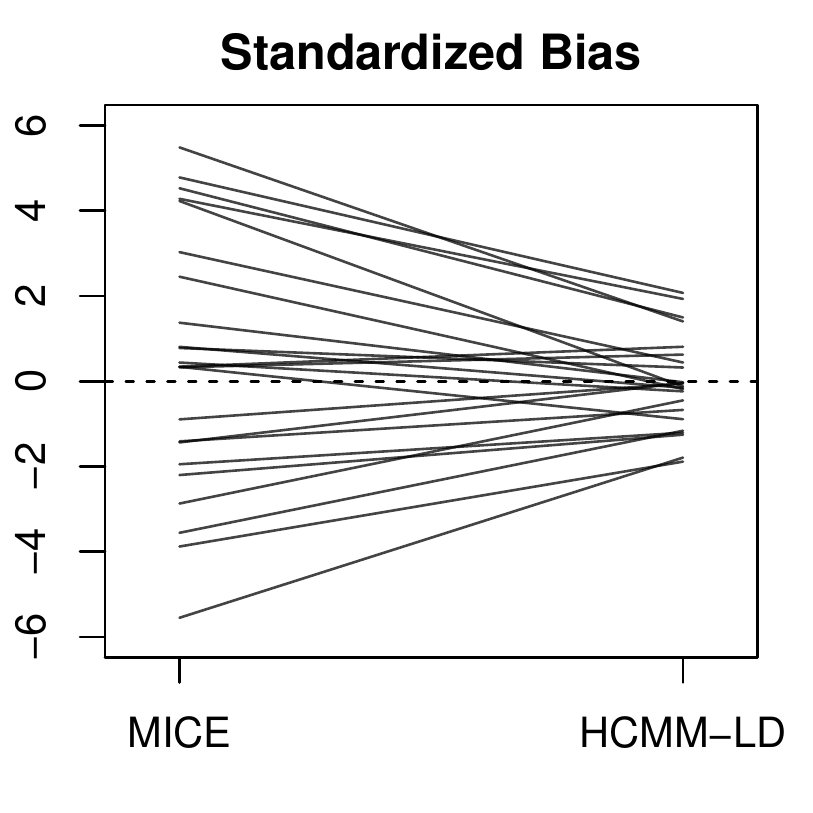} 
\includegraphics[width=.35\linewidth]{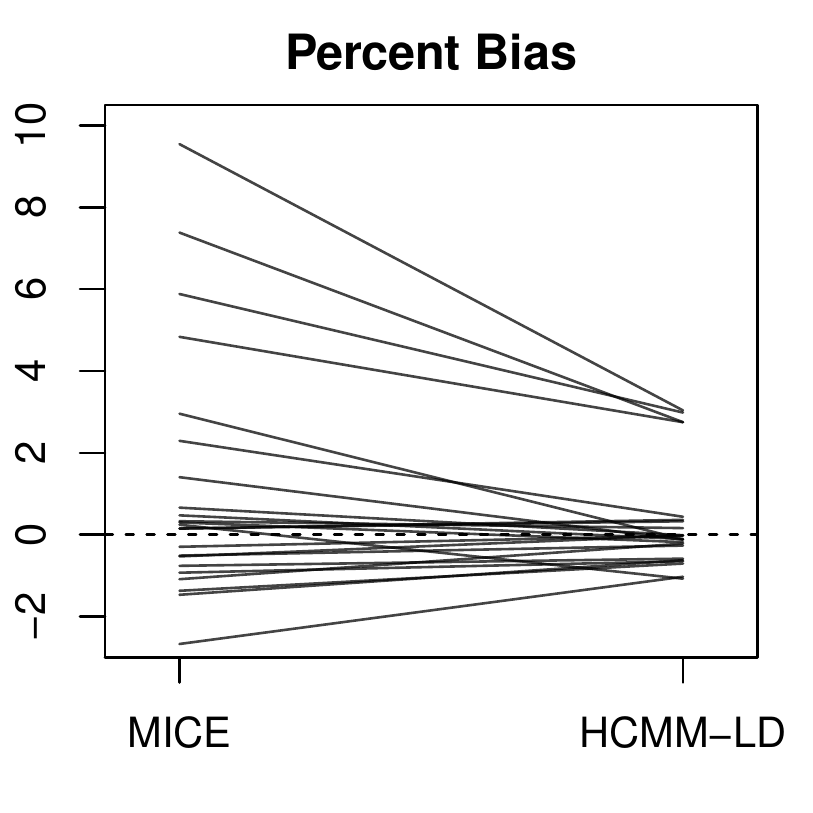} 

}

\caption{Standardized and percent bias of pooled
  estimates of mean log monthly earnings by age, sex, and own children
  in the home (Yes/No). Each line represents a cell mean, with the
  left and right endpoints at the bias under MICE and \ac,
  respectively.}
\label{sfig:age-sex-kid-bias}
\end{figure}

Figure \ref{sfig:age-sex-kid-ci} shows the coverage rates and average
width of 95\% multiple imputation confidence intervals. We have labeled the same estimands as in the main paper.  The \ac\
imputations clearly have superior repeated sampling properties.  About half of MICE's intervals have
coverage under 75\%, with many under 25\% and some approaching
0\%. In contrast, the worst coverage rate with the \ac\ is just
under 75\% with the majority near or greater than the nominal 95\%
rate. When the \ac\ imputations undercover, the MICE imputations
undercover to an even greater extent. The widths of the confidence
intervals are comparable, and there are a number of instances where the
\ac\ coverage rates are larger with shorter intervals.  This suggests
that the lack of coverage in the MICE imputations is due to bias,
which is confirmed by Figure  \ref{sfig:age-sex-kid-bias}. Overall, the
range of bias under the \ac\ is much smaller than that for
MICE. As noted in Section 4 of the main text, complex relationships between age, income and the presence of children seem to be driving these results. Both methods struggle more here than under MAR, because age and sex are subject to missingness here. The \ac\ outperforms MICE by an even wider margin in this case.

\subsubsection{Regression Coefficients}

\begin{figure}

{\centering \includegraphics[width=.4\linewidth]{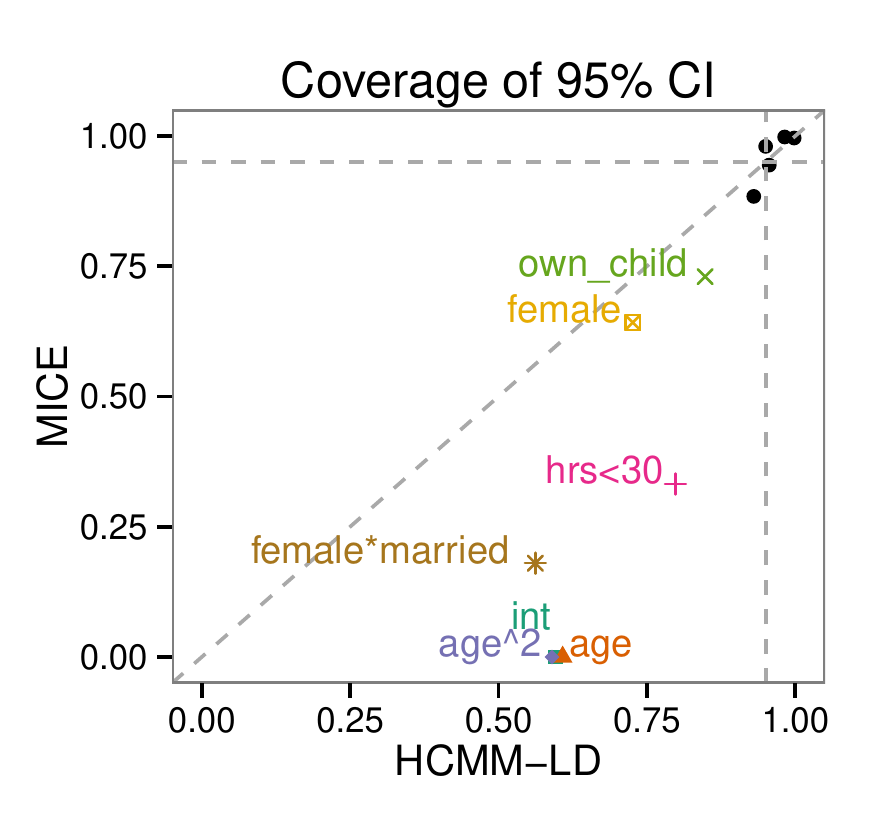} 
\includegraphics[width=.4\linewidth]{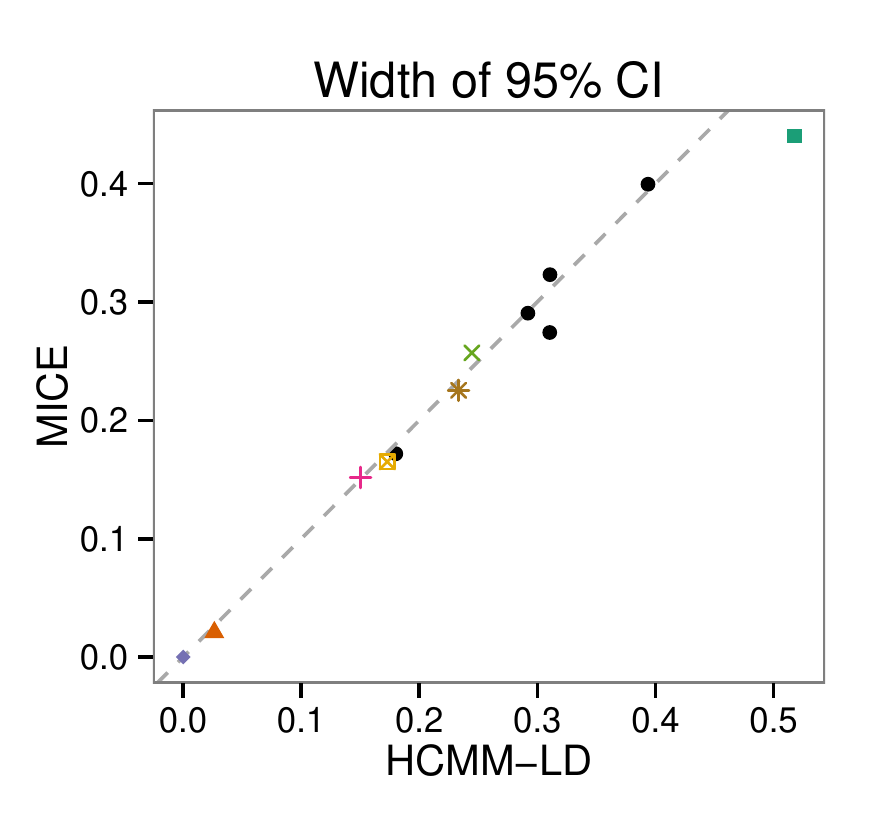} 

}

\caption{(Left) Coverage rate of pooled nominal 95\% CI for the regression with three-way interaction and age squared, including fpc. (Right) Average width
of 95\% CI.}
\label{sfig:regms3sq}
\end{figure}

Next we consider linear regressions of log earnings on age, sex, usual
hours worked (recoded as $<30$, 30-60, and $60+$), and indicators for
marriage and own child under 18 in the
household. We first fit a model including an age squared term as
well as two- and three-way interactions between sex, own child, and
marital status. Figure \ref{sfig:regms3sq} displays MI
estimates of the coefficients and the average width of their
confidence intervals. The \ac\ imputations again result in better
repeated sampling properties. Including the squared term in age is challenging for both methods, since it tends to give high leverage to points at low and high age values and neither method has been modified to anticipate the nonlinear relationship. Nonetheless, the \ac\ still offers over 50\% coverage rates for the age coefficients and the intercept, whereas the coverage rate under MICE drops to zero. The dramatic differences between Section 4.1.2 in the main paper and the results here are primarily due to the additional missingess under MCAR. Again the \ac\ proves more robust than MICE to missingness in the additional variables.

\begin{figure}

{\centering \includegraphics[width=.4\linewidth]{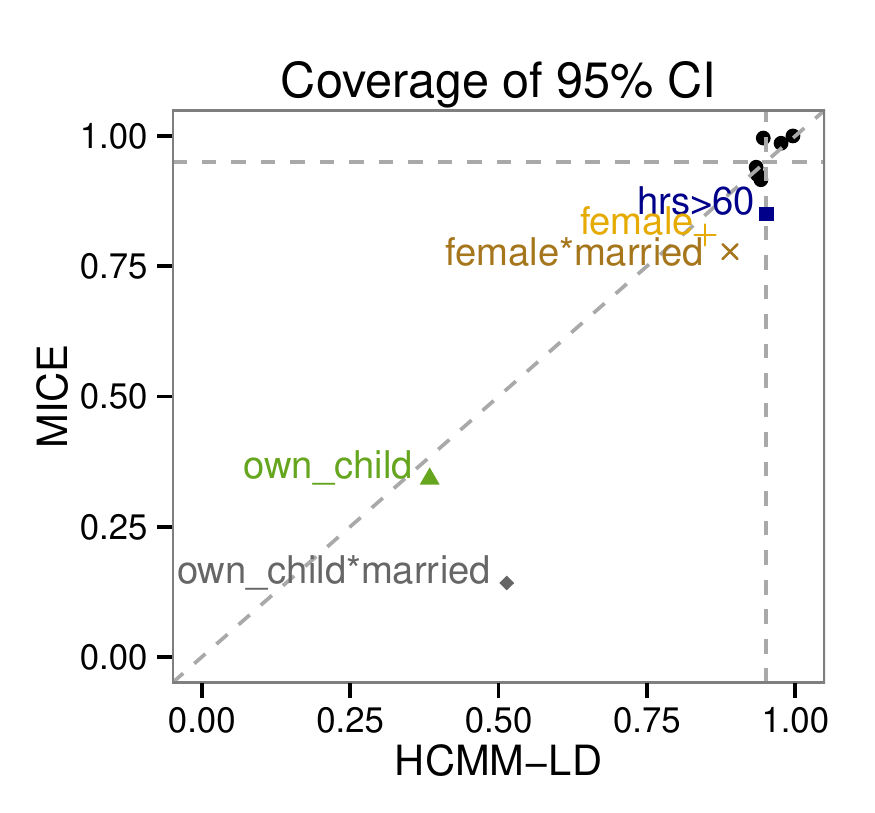} 
\includegraphics[width=.4\linewidth]{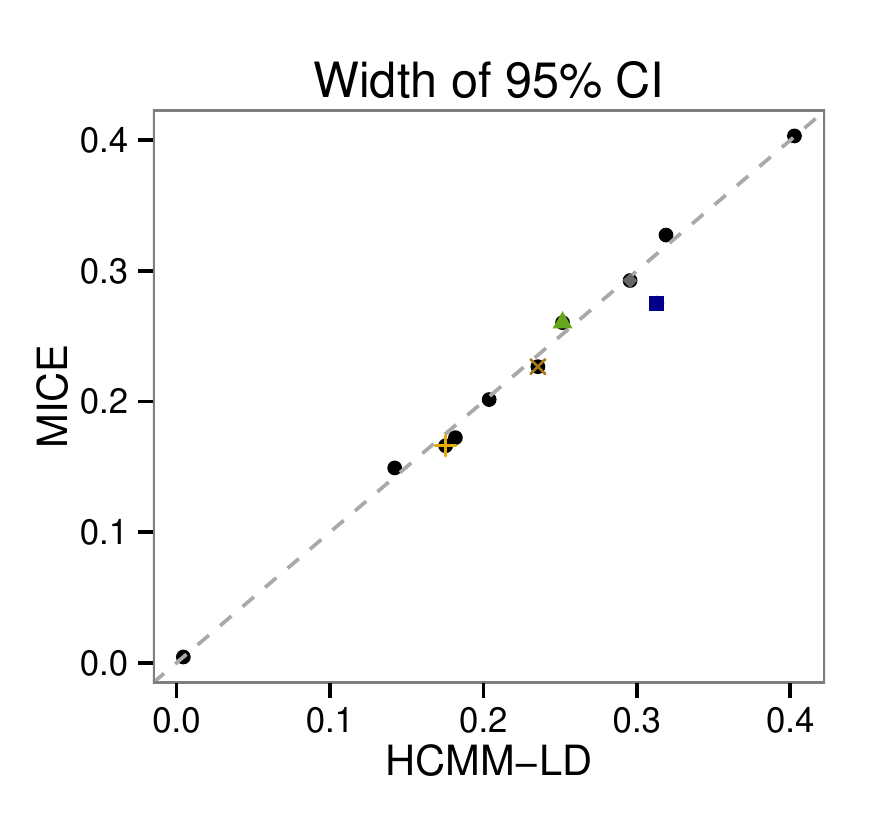} 

}

\caption{(Left) Coverage rate of pooled nominal 95\% CI for regression with three-way interaction \emph{without} age squared, including fpc. (Right) Average width
of 95\% CI.}
\label{sfig:regms3}
\end{figure}

Figure \ref{sfig:regms3} shows results from the same 3-way regression
model without the age squared term. Coverage is generally improved for
both methods.  The \ac\ imputations tend to yield moderately better coverage
rates, particularly for the two way interactions. Under both methods
the interactions are pulled toward zero, but more so with MICE
compared to the \ac. %

Considering the regression with main effects only, we find that on most coefficients the \ac\ and mice have similar properties and have roughly the nominal coverage rate. However, as in Section 4.1.2 we find that the mice imputations yield CIs for the coefficient for the indicator of usual hours worked $>60$ with coverage of about 80\%, compared to 90\% under the \ac. Again, this seems to be a problem with predictive mean matching (see Section 4.1.2 for more discussion).

\subsubsection{Conditional Frequencies}

\begin{figure}
{\centering \includegraphics[width=.4\linewidth]{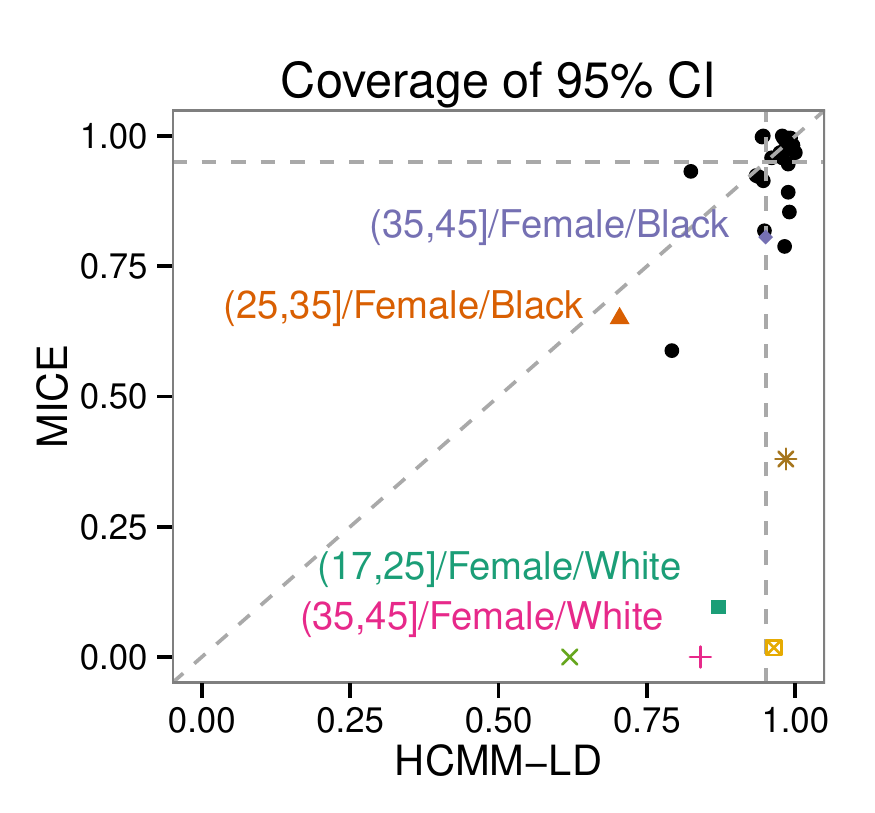} 
\includegraphics[width=.4\linewidth]{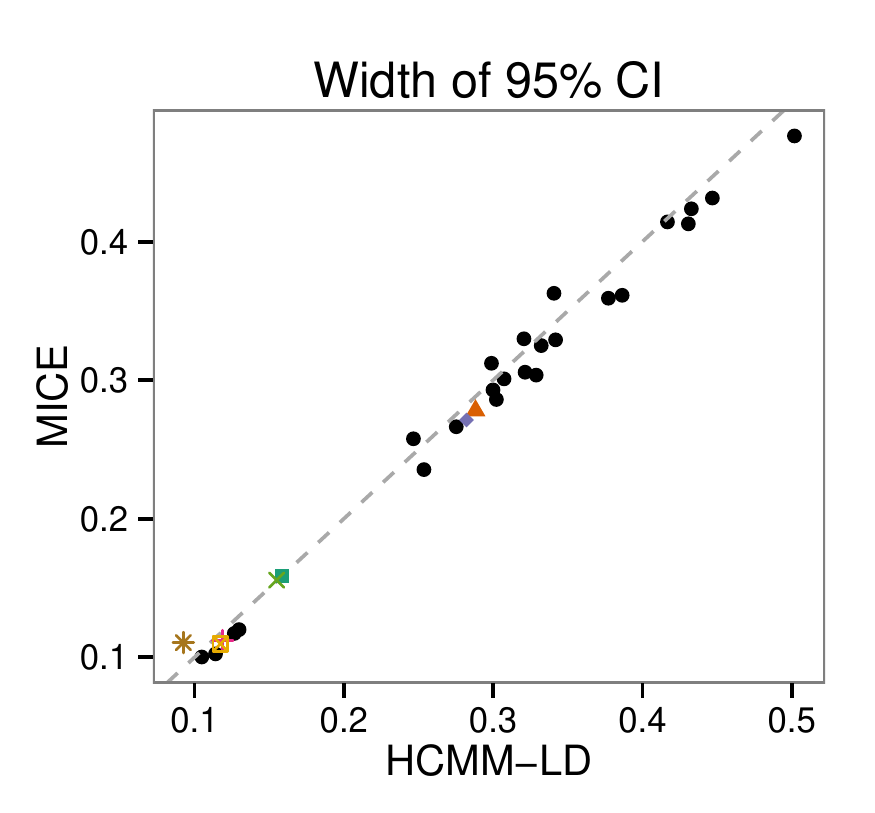} 

}
\caption{(Left) Coverage rate of nominal 95\% CIs for proportion with
  own child $<18$ in the household by age, race and sex. (Right) Average width
of 95\% CI.}
\label{sfig:prop_own_child}
\end{figure}

\begin{figure}
{\centering \includegraphics[width=.65\linewidth]{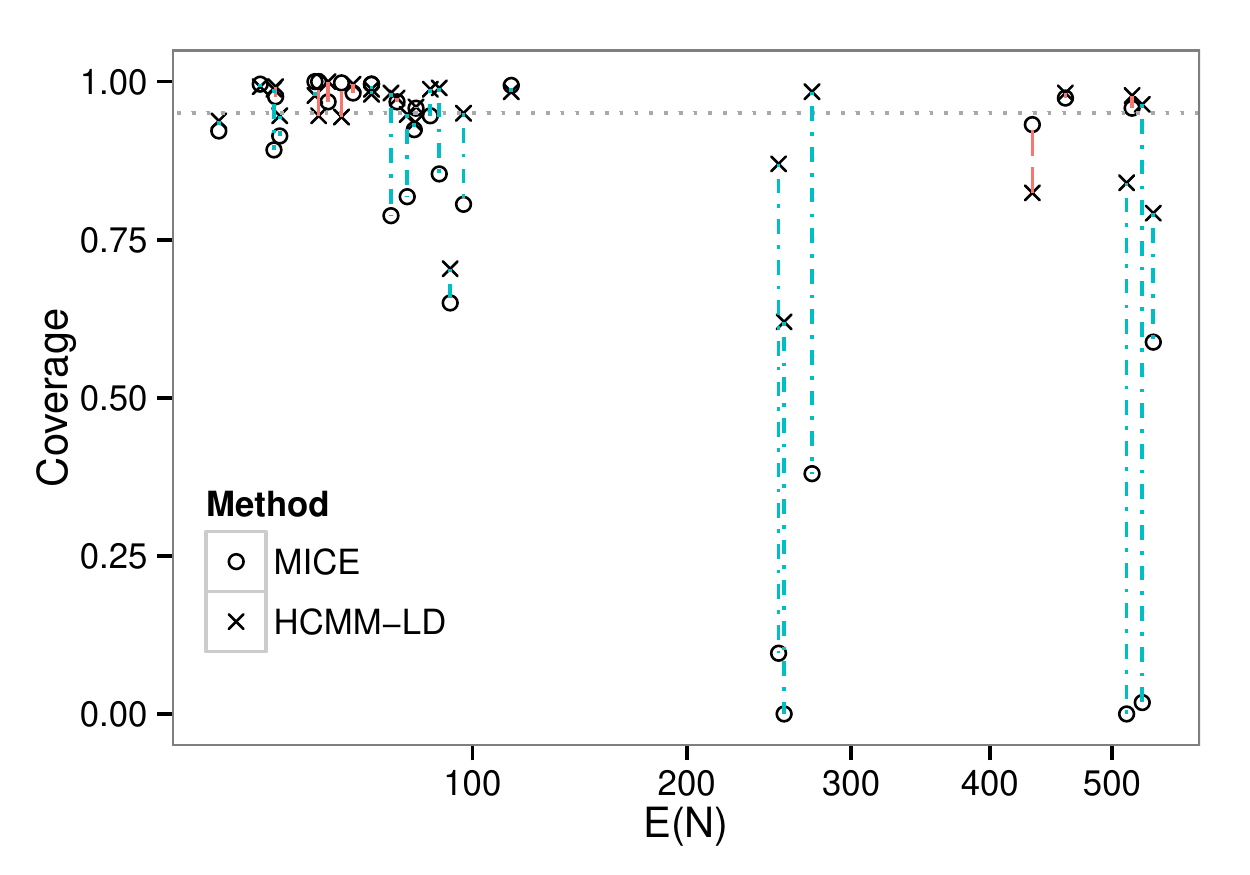} 

}
\caption{Coverage by expected cell size for proportion with own child $<18$ in the household by age, race and sex.}
\label{sfig:prop_own_child_cov_by_n}
\end{figure}

Finally we consider conditional frequencies as in Section 4.1.3 of the main text. Figure \ref{sfig:prop_own_child} displays results from estimating the proportion of respondents with their own child under 18 in the home by sex, race and age. We have labeled the same estimands here as we did in the main paper. 
The \ac\ based imputations perform better than MICE, for which some
coverage rates drop all the way to zero. Coverage rates for the \ac\
never drop below 60\% (versus 0\% for mice on that estimand) and are greater than those for MICE in every
case but one. Figure \ref{sfig:prop_own_child_cov_by_n} shows that mice
has very good or very poor coverage in large cells, consistent with
the lack of coverage arising from misspecification bias. The \ac\
tends to have somewhat lower coverage in these larger cells than in the smaller cells, but not nearly to the extent of MICE. This is probably due to finite-sample bias; larger cells are more sensitive to finite sample bias since the complete data standard errors are smaller. This effect should improve in large samples. 

\section{MCMC Sampling for the \ac}\label{supp3}

\input{supplementbody}

\bibliographystyle{./plainnat}
 \bibliography{./2013-08-mixedimp}

\end{document}

%% file: supplementbody.tex

We draw samples from the posterior via a Gibbs sampling algorithm, which we
outline below. \cite{Banerjee2013} describe an exact partially
collapsed Gibbs sampler for ITF mixtures with $\zmax,\ \kx,\ \ky$ all equal to $\infty$ which could be adopted
directly. However, in the paper we use a truncation approximation which is
simpler to implement and quite accurate, building on
\cite{Ishwaran2001}'s blocked Gibbs sampler for truncated stick
breaking priors. Below, $X^{mis}_i$ and $X^{obs}_i$ refer to the subvectors of $X_i$ which are missing or observed, respectively, with $Y^{mis}_i$ and $Y^{obs}_i$ defined similarly.  

\begin{itemize}
 \item $Z$: For each observation, sample $Z_i$ from 
\[
 \Pr(Z_i=z\mid (\Hxi, \Hyi)=(\hx, \hy), -) \propto \lambda_z\byx{\phi_{z\hx}}\byy{\phi_{z\hy}}
\]
for $1\leq z\leq \zmax$

\item $X^{mis}$: For each observation $i$ sample each missing entry of $X_i$ from its full conditional distribution
\[
 \Pr(X_{ij}=c_j\mid (\Hxi, \Hyi)=(\hx, \hy), X_{i/j}, -) \propto \psi_{\hx c_j}^{(j)}N(Y_i;\ D(\tilde x(j, c_j))B_{\hy }, \Sigma_{\hy })
\]
where $X_{i/j}$ is the $X-$vector with the $j^{th}$ element removed, and $\tilde x(j, c_j)$ is the vector with entries equal to $X_{il}$ for $l\neq j$ and $c_j$ for $l=j$.
If the number of categorical variables subject to missingness is relatively small then it may be feasible to update all the missing entries in $X_i$ in a block, which will lead to a better mixing chain when there are strong dependencies in the distribution of $X$. In practice we find this simpler update to work quite well, and is much more efficient computationally as the state space for the block update gets large rapidly as the number of missing variables increases. 

\item $\Hx$: For each observation update $\Hxi$ from 
\[
 \Pr(\Hxi=\hx\mid Z_i=z, X_i=x_i,-) \propto \byx{\phi_{z\hx}} \prod_{j=1}^{p}\psi_{\hx{x_{ij}}}^{(j)}
\]

\item $(Y^{mis}, \Hy)$: Update the cluster index for $Y$ and the missing entries in a block, by first sampling $\Hyi$ marginally over $Y^{mis}_i$ according to 
\[
 \Pr(H_{yi}=\hy\mid Z_i=z, Y_i^{obs}=y_i^{obs}, X_i=x_i, \byy{\phi_z}, \{B_h, \Sigma_h\}) \propto \phi^{(y)}_{z\hy } N(y^{obs}_i;\ D(x_i)B^{*}_{\hy}, \Sigma^{*}_{\hy})
\]
where $B^{*}_{\hy }$ is obtained by dropping the columns of $B_{\hy }$ corresponding to missing observations in $Y_i$ and $\Sigma^{*}_{\hy }$ is the relevant submatrix of $\Sigma_{\hy }$. Given the new cluster index sample the missing entries of $Y$ from 
\[
 (Y_i^{mis}\mid \Hyi=\hy -)\sim N(\tilde \mu + D(x_i) \tilde B_{\hy}, \tilde \Sigma_{\hy})
\]

where $\tilde B_\hy$ is the submatrix of the coefficients corresponding to $Y_i^{mis}$ (i.e., the columns dropped from $B_\hy$ to obtain $B^*_\hy$) and $\tilde\mu_\hy,\ \tilde\Sigma_\hy$ are available through standard calculations, since 
\[
((Y_i^{obs}, Y_i^{mis})\mid \Hyi=\hy, X_i=x) \sim N(D(x_i)B_\hy, \Sigma_\hy)
\]
after suitably permuting the rows of $B_\hy$ and the rows and columns of $\Sigma_\hy$.

\item Component parameters: For each $1\leq z\leq \zmax$, $1\leq \hx\leq \kx$ and $1\leq j\leq p$ sample
\[
 (\psi_{\hx}^{(j)}\mid -) \sim Dir\left(\gamma_{\hx 1} + \sum_{i=1}^n \ind{\Hxi=\hx, X_{ij}=1},\dots,\gamma_{\hx c_j} + \sum_{i=1}^n \ind{\Hxi=\hx, X_{ij}=d_j}\right)
\]

For each  $1\leq \hy\leq \ky$ and $1\leq v\leq q$ let $B_{\hy v}$ be the $v^{th}$ column of $B_\hy$, and sample
\[
 (B_{\hy v}\mid -)\sim 
 N\left( 
  (\tau_v I + \mathbf{D}_{\hy}'\mathbf{D}_{\hy}/\tilde\sigma_{\hy v}^2)^{-1}(\tau_v B_{0v} + \mathbf{D}_{\hy}'\mathbf{\tilde y}_{\hy v}/\tilde\sigma^2_{\hy v} ), 
  (\tau_v I + \mathbf{D}_{\hy}'\mathbf{D}_{\hy}/\tilde\sigma_{\hy v}^2)^{-1}
\right)
\]
where $\mathbf{D}_{\hy}$ is the matrix obtained by stacking the vectors $\{D(x_i): \Hyi = \hy\}$, $\mathbf{\tilde y}_{\hy v}$ is the vector obtained by concatenating $\{y_{iv}-\tilde\mu_{iv} : \Hyi=\hy\}$, and $\tilde\mu_{iv}, \tilde\sigma^2_{\hy v}$ are parameters of the conditional normal distribution for $(Y_{iv}\mid Y_{i/v}, \Hyi=\hy,-)$, where

\[
(Y_{iv}\mid Y_{i/v}, \Hyi=\hy,-)\sim N(D(X_i)B_{rv} + \tilde\mu_{iv}, \tilde\sigma^2_{\hy v})
\]

Finally, for each $1\leq \hy\leq \ky$ sample

\[
 \Sigma_{\hy}\sim IW(d + \sum_{i=1}^n \ind{\Hyi=\hy}, \Sigma + S_{\hy})
\]
where $S_{\hy} = \sum_{i:\Hyi = \hy} (Y_i-D(x_i)B_{\hy})(Y_i-D(x_i)B_{\hy})'$

\item Hyperparameters: For each entry of $B_0$ sample
\[
 (B_{0jv}\mid -) \sim N\left((\ky\tau_v+1/\sigma_0^2)^{-1} \tau_v\sum_{\hy=1}^{\hy} B_{\hy jv}, (\ky\tau_v+1/\sigma_{0\beta}^2)^{-1}\right)
\]

For $1\leq v\leq q$ sample 
\[
(\tau_v\mid -)\sim G\left( \frac{a_{\tau}+k_yp^*}{2}, \frac{b_{\tau} + \sum_{\hy=1}^{\ky} (B_{\hy v}-B_{0v})'(B_{\hy v}-B_{0v})}{2}\right)
\]

\item Mixing proportions: Resample $\lambda$ by sampling (for $1\leq z \leq \zmax-1$)
\[
(\byz{\xi}_{z}\mid -)\sim Beta\left (1 + m_z, \alpha + n - \sum_{l=1}^{z} m_l\right)
\]
where $m_z=\sum_{i=1}^n \ind{Z_i=z}$, and set $\lambda_z = \byz{\xi}_{z}\prod_{l<h}(1-\byz{\xi}_{z})$. 

For $1\leq z\leq \zmax$, iterate over $1\leq \hx\leq \kx-1$ and $1\leq \hy\leq \ky-1$ sampling
\begin{align*}
(\byx{\xi}_{z\hx}\mid -)\sim Beta\left (1 + \byx{t}_{z\hx}, \byx{\beta} + m_z - \sum_{l=1}^{\hx} \byx{t}_{zl}\right)\\
(\byy{\xi}_{z\hy}\mid -)\sim Beta\left (1 + \byy{t}_{z\hy}, \byy{\beta} + m_z - \sum_{l=1}^{\hy} \byy{t}_{zl}\right)
\end{align*}
where $\byx{t}_{z\hx} = \sum_{i=1}^n \ind{Z_i=z}\ind{\Hyi=\hy}$ and $\byy{t}_{z\hy} $ is defined similarly. Set 
\[
\byx{\phi}_{z\hx}=\byx{\xi}_{z\hx}\prod_{l<h}(1-\byx{\xi}_{zl})
\]
\[
\byy{\phi}_{z\hy}=\byy{\xi}_{z\hy}\prod_{l<h}(1-\byy{\xi}_{zl})
\]

\item Concentration parameters: Let $\mathcal{Z}_{occ}$ be the set of occupied top-level clusters (those with $m_z>0$), and let $n_{occ} = |\mathcal{Z}_{occ}|$. Sample the concentration parameters from their gamma full conditionals:
\[
 \alpha\sim G(a_0+\zmax-1, b_0-\log(\lambda_{\zmax}))
 \]
 \begin{align*}
\byx{\beta} &\sim G\left(\byx{a}+n_{occ}(\kx-1), \byx{b}-\sum_{z\in\mathcal{Z}_{occ}}\log\left(\byx{\phi}_{z\kx}\right) \right)\\
\byy{\beta} &\sim G\left(\byy{a}+n_{occ}(\ky-1), \byy{b}-\sum_{z\in\mathcal{Z}_{occ}}\log\left(\byy{\phi}_{z\ky}\right) \right).
\end{align*}
In the paper we take $\byx{a}=\byy{a}=\byx{b}=\byy{b}=0.5$

\end{itemize}